\documentclass{JHEP3}
\pdfoutput=1

\usepackage{graphicx}
\usepackage{bm}
\usepackage{amsmath,amssymb}
\usepackage{comment}
\usepackage{setspace}
\usepackage{float}

\newcommand{\nt}{\nonumber\\}

\newcommand{\cN}{{\cal N}}

\newcommand{\tr}{{\rm tr}}

\newcommand{\ba}{\begin{eqnarray}}
\newcommand{\ea}{\end{eqnarray}}

\newcommand{\eps}{\epsilon}

\newcommand{\mathsym}[1]{{}}
\newcommand{\unicode}[1]{{}}

\title{M2- and M5-branes in E11 Current Algebra Formulation of M-theory}
\preprint{KEK-TH-1996}

\author{
Shotaro Shiba$^{\diamondsuit}$ and Hirotaka Sugawara$^{\spadesuit}$
\vspace*{3mm}\\
${}^\diamondsuit$
Theory Center, High Energy Accelerator Research Organization (KEK),\\
1-1 Oho, Tsukuba, Ibaraki 305-0801, Japan.\\
\email{sshiba@post.kek.jp}
\vspace*{3mm}\\
${}^\spadesuit$
Okinawa Institute of Science and Technology Graduate University (OIST),\\
1919-1 Tancha, Onna-son, Kunigami, Okinawa 904-0495, Japan.\\
\email{hirotaka.sugawara@oist.jp}
}

\abstract{
Equations of motion for M2- and M5-branes are written down in the $E_{11}$ current
algebra formulation of M-theory. 
These branes correspond to currents of the second and the fifth rank antisymmetric tensors in the $E_{11}$ representation,
whereas the electric and magnetic fields (coupled to M2- and M5-branes) correspond to
currents of the third and the sixth rank antisymmetric tensors, respectively.
We show that these equations of motion have solutions in terms of the coordinates on M2- and M5-branes. 
We also discuss the geometric equations, and show that there are static solutions 
when M2- or M5-brane exists alone and also when M5-brane wraps around M2-brane.
This situation is realized because our Einstein-like equation contains an extra term
which can be interpreted as gravitational energy contributing to the curvature,
thus avoiding the usual intersection rule.
}


\begin{document}

%%%%%%%%%%%%%%%%%%%%%%%%%%%%%
\section{Introduction}
%%%%%%%%%%%%%%%%%%%%%%%%%%%%%

%\noindent
%{\bf (Brief history of studies on M2- and M5-branes)}
%\vspace{5pt} \noindent

M-theory is believed to be a nonperturbative description of superstring theories.
Many researchers have been studying this theory to clarify various aspects of superstrings.
M-theory is reduced to 11d supergravity in the low energy limit. 
This supergravity has solutions of black branes,
which have electric or magnetic charges as well as energy-momentum
in 11d spacetime.
% can be a source of both gravitational and elecromagnetic fields
In M-theory, % or in high energy scale,
these electric and magnetic charges of black branes are thought to be quantized.
A brane with a single electric charge is called M2-brane,
while a brane with a single magnetic charge is M5-brane.
Moreover, these M-branes are dynamical objects in 11d spacetime
and are considered to play the central role in M-theory.
Up to now, many attempts have been done to describe their behaviors,
especially, in terms of field theory defined on the brane worldvolume.

A field theory on a single M2-brane was formulated in 1980's~\cite{Bergshoeff:1987cm,deWit:1988wri}.
A theory on multiple M2-branes was firstly proposed as BLG theory,
where gauge symmetry is described using Lie 3-algebra~\cite{Bagger:2007jr,Gustavsson:2007vu}.
Soon after that, ABJM theory with $U(N)\times U(N)$ gauge symmetry 
was proposed to describe a theory on $N$ M2-branes~\cite{Aharony:2008ug}.
In particular, the free energy of this ABJM theory can be formulated in terms of matrix model~\cite{Kapustin:2009kz,Drukker:2010nc}.
One of the authors (S.S.) have analyzed BLG theory and ABJM matrix theory, 
and clarified some dynamical aspects of M2-branes~\cite{Ho:2008ve,Chu:2008qv,Ho:2009nk,Kobo:2009gz,Hanada:2012si}.

Compared with M2-brane, a field theory on M5-brane is difficult to formulate
due to the self-duality of the 2-form field on the brane.
A theory on a single M5-brane was formulated using a nontrivial auxiliary field~\cite{Pasti:1997gx},
but at this moment 
we have no consensus about the theory on multiple M5-branes.
Some researchers have proposed that it may be described using Lie 3-algebra~\cite{Lambert:2010wm,Honma:2011br},
an algebra including nonlocal operators~\cite{Ho:2011ni,Huang:2012tu}
or more exotic algebra~\cite{Samtleben:2011fj}.
On the other hand, some researchers suggested that
all information on multiple M5-branes may be contained in a field theory on multiple D4-branes in superstring theory~\cite{Douglas:2010iu,Lambert:2010iw}.
In spite of many attempts, we have not obtained a satisfactory formulation.

Here we would like to propose a new approach to study M-brane dynamics.
Our approach is based on a formulation of M-theory in terms of $E_{11}$ algebra proposed by P. West~\cite{West:2003fc} and his collaborators,
and the current algebra formulation by one of the authors (H.S.)~\cite{Sugawara:2017fds}.

%\vspace{10pt} \noindent
%{\bf (On the $E_{11}$ current algebra formulation of M-theory) by Sugawara-san}

In a series of papers with his coworkers, P.~West studied M-theory based on the nonlinear representation of the $E_{11}$ Kac-Moody algebra.
There are subsequent studies of the issue by other authors~\cite{Englert:2008ft, Berman:2014hna}. 
To quantize this theory, one of the authors (H.S.) adopted the current algebra method~\cite{Sugawara:1967rw} 
rather than the usual canonical or the path integral method. 
The idea is to use the current algebra commutation relations rather than the
canonical commutation relations. 
The energy-momentum tensor can be written in terms of bilinear form of the currents,
and the quantum equation of motion can be derived simply from
\ba
-i\partial_\mu J_\nu = [P_\mu, J_\nu]
\ea
where $J_{\nu}$ can be any currents that appear in $E_{11}$ representations and
\ba
P_{\mu} = \frac{1}{C}\int \Theta_{\mu 0}(x) dx 
= - \frac{1}{C}\int \left( J_{\mu} J_0 - \frac{1}{2}\eta_{\mu 0} J^{\nu }J_{\nu }\right) dx\,.
\ea

The currents of the $E_{11}$ include elfbein and spin connections that appear in the gravity theory, in addition to the various antisymmetric representations of $E_{11}$ algebra. 
This is made possible by using the graded algebra of $E_{11}$, which means that we use not only the adjoint representation of the $E_{11}$ but other representations to define the currents.

We now argue, or, rather, explain our motivation of why we use Kac-Moody algebra and why specifically the $E_{11}$ algebra. 
We then argue why we think it is appropriate to use the current algebra method to quantize the theory.

First of all, it is well known by now that the $SU(N)$ gauge theory with $N \rightarrow \infty$ describes the string. 
And, if the string includes the closed one, $SU(\infty)$ gauge theory has the potential to describe the gravity. 
This argument can be further justified when we understand the relation between the $SU(\infty)$ algebra and the diffeomorphism~\cite{Pope:1989cr} 
that plays an important role in the gravity theory. 
There is no direct proof of %We still do not understand 
the relation between the $SU(\infty)$ algebra and the Kac-Moody algebra, 
but there already exist some works which investigate the relation between the Kac-Moody algebra and the diffeomorphism~\cite{Frappat:1989gn}
implying indirectly the relation between $SU(\infty)$ and the Kac-Moody algebra. 
This suggests the possible relation between the $SU(\infty)$ algebra and the Kac-Moody algebra. 
This motives our use of Kac-Moody algebra rather than the $SU(\infty)$ algebra in describing the strings and especially the gravity. 

The next question is: why specifically $E_{11}$ algebra?
There is a strong indication that the 10d supergravity theories have $E_8$
symmetry~\cite{Brink:2008hv}. 
Since we are aware that the string theory must include 10d supergravity, our Kac-Moody algebra must include $E_8$ algebra as its subalgebra. 
The extended Kac-Moody algebra $E_9$ will correspond to the inconsistent string theory lacking the Liouville mode. 
Adding the Liouville mode gives rise to the {``}very extended{''} Kac-Moody algebra $E_{10}$, and this will correspond to the consistent string theory. 
The natural next step is to go to the {``}over extended{''} algebra $E_{11}$ to describe M-theory and this was extensively investigated by P. West and his collaborators~\cite{West:2003fc}.
It is possible that the consistent F-theory may be formulated by using the $E_{12}$ Kac-Moody algebra, and it will be one of our future targets. 
Fig 1. shows the situation described here.

\FIGURE{
%\begin{figure}[H]
%\centering
\includegraphics[width=12cm]{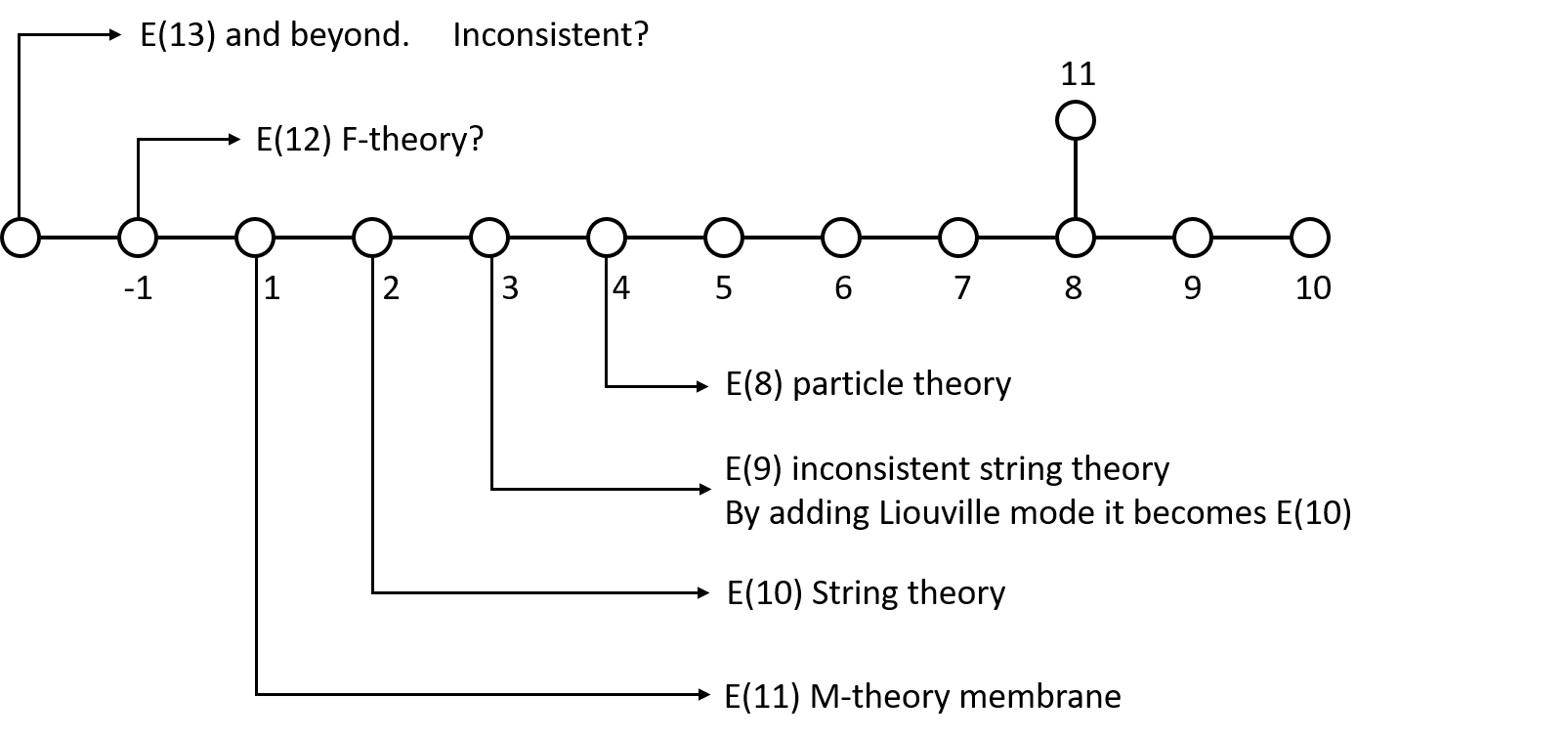}
\caption{Role of extended, very extended and over extended algebras}
%\end{figure}
}

The next question is: why current algebra formulation rather than the usual quantum theory formalism with symmetry? 
To answer this question, we refer to the work done in 1968 by Bardacki, Frishman and Halpern~\cite{Bardakci:1968zz}.

First, remember that, to get the string theory from $SU(N)$, we need
\ba
N\rightarrow \infty
\quad \text{with}\quad
g^2N~\text{fixed}.
\ea
On the other hand, 
Bardacki, Frishman and Halpern proved that the massive Yang-Mills theory becomes current algebra theory with current-current energy-momentum tensor if we take the limit:
\ba
g\rightarrow 0 \quad\text{and}\quad
m\rightarrow 0\,.
\ea
This implies that large $N$ limit of massive $SU(N)$ gauge theory becomes the current-current theory, if we take
\ba
m = \frac{m_0}{N}\,,\quad
N\rightarrow \infty \quad
\text{with}\quad
m_0 \text{ and } g^2N \text{ fixed}.
\ea
Therefore, the $N \rightarrow \infty$ limit of $SU(N)$ gauge theory is in fact the current-current theory presumably with some kind of Kac-Moody symmetry.

This concludes the explanation of our use of $E_{11}$ Kac-Moody current-current theory to describe M-theory.

Next, we describe what kinds of currents we use in the following sections to study M2- and M5-branes. 
All these currents are in the representation of $E_{11}$ algebra. 
In fact, we use the graded algebra $E_{11} \otimes \ell_1$, where $\ell_1$ is the ``vector representation'' %in level 1 
of $E_{11}$ algebra~\cite{West:2003fc}. 
We can also include a spinor representation of $E_{11}$ to make the theory supersymmetric~\cite{Sugawara:2017fds}.

\vspace{7pt}
\noindent
(1) We have geometric currents $k_{\mu }^{ab}(x)$ and $e_{\mu }^a(x)$:
the former belongs to the adjoint representation of $E_{11}$ and 
the latter to the $\ell_1$ representation. 
$k_{\mu }^{ab}(x)$ is related to the spin connection $\omega_{\mu }^{ab}(x)$ by
\ba
\omega _{\mu }^{ab}(x)=\frac{i}{2C}k_{\mu }^{ab}(x)\,.
\ea
By convention, we assume $k_\mu^{ab}(x)$, $e_\mu^a(x)$ and other currents defined below to be antihermitian. 
Therefore, the spin connection defined above is hermitian,
and we also have $-ie_{\mu }^a(x)$ as the hermitian elfbein. 

These two currents $k_\mu^{ab}(x)$ and $e_\mu^a(x)$ should describe the gravity theory.
However, since we get their equations of motion from
$-i\partial _{\mu }J_{\nu }=\left[P_{\mu }, J_{\nu }\right]$
with $J_\mu=$ either $k_{\mu}^{ab}(x)$ or $e_\mu^a(x)$, 
there is no guarantee that we get the Einstein equation. 
In fact, our equation deviates from the Einstein equation in a significant way,
as we will see.
To make the theory supersymmetric, we introduce the supergravity field 
$\psi _{\mu }(x)$ which plays the role of supercurrent, 
that is, the space integral of this field is nothing but the supersymmetry operator. 
This is an example of field-current identity of our theory~\cite{Sugawara:2017fds}. 
The space integral of the elfbein field $e_\mu^a(x)$ is the ``energy-momentum'' in the tangent space.

\vspace{7pt}
\noindent
(2) To describe M2- and M5-branes, we need the ``brane'' currents %-field 
$W_{\mu }^{ab}(x)$ and $W_{\mu }^{abcde}(x)$:
both belong to the $\ell_1$ representation. 
$W_{\mu }^{ab}(x)$ is the second rank antisymmetric tensor and
$W_{\mu }^{abcde}(x)$ is the fifth rank antisymmetric tensor. 
In addition, we need {``}brane charge{''} currents $B_{\mu }^{abc}(x)$ and $B_{\mu }^{abcdef}(x)$
to describe the electric and magnetic fields coupled to M2- and M5-branes.\footnote{
To be precise, 
the currents $B_\mu^{abc}(x)$ and $B_\mu^{abcdef}(x)$ are related to 
the 3-form field $C_{\mu\nu\rho}(x)$ and its dual 6-form field $C_{\mu\nu\rho\sigma\tau\lambda}(x)$ in 11d supergravity through the non-linear realization. 
(See Appendix \ref{sec:ca}.)}
These currents belong to the adjoint representation of $E_{11}$, and 
play the role of sources of M2- and M5-brane currents in the respective equations for 
$W_{\mu }^{ab}(x)$ and $W_{\mu }^{abcde}(x)$.
$B_{\mu }^{abc}(x)$ and $B_{\mu }^{abcdef}(x)$ also have their own equations to be satisfied, and 
$W_{\mu }^{ab}(x)$ and $W_{\mu }^{abcde}(x)$ play the role of sources to these equations in return. % reverse?

\vspace{7pt}
We will show in the following sections that these equations can be satisfied by certain expressions in terms of the coordinates on M2- or M5-branes.

%%%%%%%%%%%%%%%%%%%%%%%%%%%%
\section{$E_{11}$ current algebra}
\label{sec:E11}
%%%%%%%%%%%%%%%%%%%%%%%%%%%%

\subsection{Formulation}

Let us first define the vector currents $J_\mu^A(x)$ in our theory as
\ba \label{Omega}
\Omega_\mu(x)
&=& J_\mu^A(x) G^A \nt
&=& k_\mu^{ab} (x) K_{ab} %k_\mu{}^b{}_a (x) K^a{}_b 
+ e_\mu^a(x) P_a + B_\mu^{abc}(x) S_{abc} 
+B_\mu^{abcdef}(x) S_{abcdef} \nt 
%+ \psi^\alpha_\mu(x) \Psi_\alpha \nt
&&
 +\, W_\mu^{ab}(x) Z_{ab} + W_\mu^{abcde}(x) Z_{abcde} + W_{\mu}^{abcdefg,h}(x) Z_{abcdefg,h}
 + \cdots
\ea
where $G^A$ are generators of the $E_{11}\otimes \ell_1$ algebra.
The index $A$ denotes all the independent elements of the generators.
The indices $a, b, \ldots = 0, \ldots, 10$ of the generators and currents are 
totally antisymmetric, as mentioned in Introduction,
and they can be raised or lowered by $SO(10,1)$ Lorentzian metric $\eta_{ab}$. 
The index $\mu=0,\ldots,10$ denotes all the directions in 11d curved spacetime,
which is raised or lowered by the metric $g_{\mu\nu}(x) = \eta_{ab} e_\mu^a(x) e_\nu^b(x)$.

The generators of $E_{11}$ algebra are often classified by its level: 
the level of a root of $E_{11}$ algebra is defined to be a multiplicity 
of a component of a specific simple root in its decomposition into simple roots. (See the references~\cite{West:2003fc,Sugawara:2017fds} for details.)
For example, $E_{11}$ algebra contains the generators 
$\tilde K^a{}_b$, $R^{abc}$ and $R^{abcdef}$
at level 0, 1 and 2, respectively.
Similarly, the generators $R_{abc}$ and $R_{abcdef}$ % and $R_{abcdefgh,i}$
are contained at level $-1$ and $-2$. % and $-3$. 
Note that the generators $K_{ab}, S_{abc}$ and $S_{abcdef}$ in eq.\,(\ref{Omega})
are defined as 
\ba
K_{ab} &:=& 2 \eta_{c[a} \tilde K^c{}_{b]} \nt
S_{abc} &:=& R^{def}\eta_{ad}\eta_{be}\eta_{cf} - R_{abc} \nt
S_{abcdef} &:=& R^{ghijkl}\eta_{ag}\ldots\eta_{lf} + R_{abcdef}\,,
%S_{abcdefgh,i} &:=& R^{jklmnopq,r}\eta_{aj}\ldots\eta_{ir} - R_{abcdefgh,i}
\ea
% and $R^{abcdefgh,i}$
so that they are invariant under Cartan involution~\cite{West:2003fc}.
The generators $P_a, Z_{ab}, Z_{abcde}$ and $Z_{abcdefg,h} := \epsilon_{abcdefg}{}^{ijkl} Z_{hijkl}$ are in the $\ell_1$ representation with the highest weight of $E_{11}$ algebra.
The commutation relations among the generators are shown in appendix \ref{sec:comm}.

Next, according to the current algebra formulation, we define the commutation relations among the vector currents $J_\mu^A$ as 
\ba \label{current_eom}
{}[J_0^A(x), J_0^B(y)]_{x_0=y_0} &=& f^{AB}{}_C J_0^C(y)\delta(\vec x-\vec y)\nt
{}[J_0^A(x), J_N^B(y)]_{x_0=y_0} &=& f^{AB}{}_C J_N^C(y)\delta(\vec x-\vec y) 
+ iC\eta^{AB}\partial_N \delta(\vec x-\vec y)\nt
{}[J_M^A(x), J_N^B(y)]_{x_0=y_0} &=& 0
\ea
for $M,N\neq 0$. 
Here $J_\mu^A$ are antihermitian currents,
and the structure constant $f^{AB}{}_C$ is that of $E_{11}\otimes \ell_1$ algebra $[G^A, G^B] = f^{AB}{}_C G^C$.
The commutation relations for each kind of currents are shown in appendix \ref{sec:comm}.

The energy-momentum tensor is defined as a bilinear form of the currents
\ba
\Theta_{\mu\nu}(x) 
= - \frac{1}{C}\left( J^A_\mu(x) J^A_\nu(x) - \frac12 \eta_{\mu\nu} \eta^{\rho\sigma} J^A_\rho(x) J^A_\sigma(x) \right)\,,
\ea
where $C$ is a constant with the dimension of length, which can be identified with 11d Planck scale~\cite{Sugawara:2017fds}.
Here we should be careful about summations of the indices $A$.
In the case of $J_\mu^A = k_\mu^{ab}$, for example,
$J_\mu^A J_\nu^A$ means 
$\frac12 \eta_{ac} \eta_{bd} k_\mu^{ab} k_\nu^{cd}
= \frac12 k_\mu^{ab} k_{\nu,ab}$
due to antisymmetry of the indices $a,b,c,d$.
Then the  {\em quantum} equation of motions for the current can be given as
\ba
-i\partial_\mu J_\nu^A(x) = [P_\mu, J_\nu^A(x)]
\ea
where
\ba
P_\mu = \int dy \,\Theta_{0\mu}(y)
= - \frac{1}{C}\int dy\left( J^B_0(y) J^B_\mu(y) - \frac12 \eta_{0\mu} \eta^{\rho\sigma} J^B_\rho(y) J^B_\sigma(y) \right).
\ea
Note that, as a linear combination of the equations of motion, 
we always obtain the conservation law of the currents
\ba
\partial^\mu J_\mu^A (x) = 0\,.
\ea

\subsection{Equations of motion for currents}
\label{sec:EOM}

Now we can obtain the equations of motion for all the currents.
Let us list some of them in this subsection.
%the EOM's for the currents in eq.\,(\ref{Omega}).

First, the equations of motion for the brane currents $W_\mu^{ab}(x)$ and $W_\mu^{abcde}(x)$ are
\ba \label{eom_M2}
\partial_\mu W_\nu^{ab} - \partial_\nu W_\mu^{ab}
&=& \frac{i}{2C}\left(
2 k_\mu{}^{[a}{}_c W_\nu^{b]c} 
-2 B_\mu^{abc} e_{\nu,c}  
+20 B_{\mu,cde} W_\nu^{cdeab}
\right. \nt && \left. \quad
-\frac{1}{24} \eps^{abcdefg}{}_{ijkl} B_{\mu,cdefgh} W_\nu^{hijkl}
+\cdots \right)
-\left(\mu\leftrightarrow\nu\right)
\ea
for M2-brane, and
\ba \label{eom_M5}
\partial_\mu W_\nu^{abcde} - \partial_\nu W_\mu^{abcde}
&=& \frac{i}{2C}\left(
-5 k_{\mu}{}^{[a}{}_f W_{\nu}^{bcde]f}  
-2 B_{\mu}^{[abc} W_{\nu}^{de]} 
+2 B_{\mu}^{abcdef} e_{\nu,f} 
\right. \nt && \left. \quad
+\frac{1}{24} B_{\mu}^{fgh} W_{\nu}^{abcde}{}_{fg,h}
+\frac{7}{10} B_\mu^{fgh[abc} W_{\nu, fgh}{}^{de]}
+\cdots \right)
-\left(\mu\leftrightarrow\nu\right).
\nt
\ea
for M5-brane. 
Next, the equations of motion for the brane charge currents $B_\mu^{abc}(x)$ and $B_\mu^{abcdef}(x)$ are
\ba \label{eom_B3}
\partial_\mu B_\nu^{abc} - \partial_\nu B_\mu^{abc}
&=& \frac{i}{2C}\left(
-3 k_\mu{}^{[a}{}_d B_\nu^{bc]d}
+2 W_\mu^{[ab} e_\nu^{c]} 
+\frac13 B_\mu^{def} B_\nu^{abc}{}_{def} 
-60 W_\mu^{de} W_\nu^{abc}{}_{de}
\right. \nt && \left. \quad
-\frac{1}{4} W_\mu^{defgh} W_{\nu,defgh}{}^{[ab,c]} 
+\cdots \right)
-\left(\mu\leftrightarrow\nu\right)
\ea
where $W_{\mu}^{abcdefg,h}:=\epsilon^{abcdefg}{}_{ijkl}W_\mu^{{hijkl}}$
and
\ba \label{eom_B6}
\partial_\mu B_\nu^{abcdef} - \partial_\nu B_\mu^{abcdef}
&=& \frac{i}{2C}\left(
6 k_\mu{}^{[a}{}_g B_\nu^{bcdef]g}
-2 W_\mu^{[abcde} e_\nu^{f]}
-80 B_\mu^{[abc} B_\nu^{def]} 
\right. \nt && \left. \quad
+\frac{7}{10} W_\mu^{abcgh} W_\nu^{def}{}_{gh}
+15 W_\mu^{gh} W_\nu^{abc}{}_{gh}{}^{de,f} 
+\cdots \right)
-\left(\mu\leftrightarrow\nu\right).
\nt
\ea

Next, the equation of motion for the elfbein $e_\mu^a(x)$ is
\ba \label{eom_e}
\partial_\mu e_\nu^a - \partial_\nu e_\mu^a
= \frac{i}{2C}\left(
- k_\mu{}^{a}{}_b e_\nu^b 
+ B_{\mu}^{abc} W_{\nu,bc}
- B_{\mu}^{abcdef} W_{\nu,bcdef}
+ \cdots \right)
-\left(\mu\leftrightarrow\nu\right).
\ea
Here we define the covariant derivative
\ba
D_\mu{}^a{}_b := \delta^a_b \partial_\mu + \frac{i}{2C} k_\mu{}^a{}_b
\ea
such that 
\ba
D_\mu{}^a{}_b e_\nu^b - D_\nu{}^a{}_b e_\mu^b 
= \frac{i}{2C} \left(
 B_{\mu}^{abc} W_{\nu,bc}
- B_{\mu}^{abcdef} W_{\nu,bcdef}
+ \cdots \right)
-\left(\mu\leftrightarrow\nu\right).
\ea
For a general tensor $V_\mu^a$, the covariant derivative is defined as 
\ba
D_\mu V_\nu^a = \partial_\mu V_\nu^a - \Gamma^\sigma_{\mu\nu} V_\sigma^a
+ \omega_\mu{}^a{}_b V_\nu^b
\ea
where $\Gamma^\sigma_{\mu\nu}$ is the Christoffel symbol 
and $\omega_\mu{}^a{}_b$ is the spin connection:
\ba \label{omega_def}
\omega_\mu^{ab} = -\frac12 e_{\mu c} \left(\Omega^{abc} -\Omega^{bca} -\Omega^{cab}\right) 
\,,\quad
\Omega^{abc} = 2 e^{\mu a} e^{\nu b} \partial_{[\mu} e_{\nu]}^c\,.
\ea
This is in fact derived by solving the equation for the elfbein (\ref{eom_e}),
neglecting the contributions from $W_\mu$ or $B_\mu$.
Therefore, this expression (\ref{omega_def}) must be considered as an approximation. 
To get this result, we identified:
%Therefore we can identify
\ba \label{omega-k}
\omega_\mu^{ab} = \frac{i}{2C} k_\mu^{ab}\,.
\ea

Finally, the equation of motion for the connection field $k_\mu^{ab}(x)$ is
\ba \label{eom_k}
\partial_\mu k_\nu^{ab} - \partial_\nu k_\mu^{ab}
&=& \frac{i}{2C}\left(
2 k_\mu{}^{[a}{}_c k_\nu^{b]c}
+2 e_\mu^{[a} e_\nu^{b]} 
+ B_\mu^{[a}{}_{cd} B_\nu^{b]cd}
-2 W_\mu^{[a}{}_c W_\nu^{b]c} 
\right. \nt && \left. \quad
-\frac{1}{60} B_\mu^{[a}{}_{cdefg} B_\nu^{b]cdefg}
+\frac{1}{12} W_\mu^{[a}{}_{cdef} W_\nu^{b]cdef}
+\cdots \right)
-\left(\mu\leftrightarrow\nu\right).
\ea
Using eq.(\ref{omega-k}), this equation can be rewritten as 
\ba \label{eom_omega}
&&
\partial_\mu \omega_\nu^{ab} - \partial_\nu \omega_\mu^{ab}
-2 \omega_\mu{}^{a}{}_c \omega_\nu^{bc}
+2 \omega_\nu{}^{a}{}_c \omega_\mu^{bc}
\nt &&
= -\frac{1}{4C^2}\left(
2 e_\mu^{[a} e_\nu^{b]} 
+ B_\mu^{[a}{}_{cd} B_\nu^{b]cd}
-2 W_\mu^{[a}{}_c W_\nu^{b]c} 
%\right. \nt && \left. \qquad
%-\frac{1}{12} W_\mu^{[a}{}_{cdef} W_\nu^{b]cdef}
+\cdots \right)
-\left(\mu\leftrightarrow\nu\right).
\ea
Note that here this equation is different from Einstein equation, the left-hand side of which is $\partial_\mu \omega_\nu^{ab} - \partial_\nu \omega_\mu^{ab} - \omega_\mu{}^{a}{}_c \omega_\nu^{bc} + \omega_\nu{}^a{}_c\omega_\mu^{bc}$.
In addition, using the definition of Riemann tensor 
\ba \label{Riemann}
R^\lambda_{\sigma\mu\nu} 
= e_c^\lambda e_\sigma^d \left(
\partial_\mu \omega_\nu{}^c{}_d
- \partial_\nu \omega_\mu{}^c{}_d
+ \omega_\mu{}^c{}_b \omega_\nu{}^b{}_d 
- \omega_\nu{}^c{}_b \omega_\mu{}^b{}_d \right),
\ea
% as we will see in \S\,\ref{sec:geometric},  this 
eq.\,(\ref{eom_omega}) leads to
\ba \label{Einstein-like}
R_{\mu\nu} - \frac{5}{C^2} g_{\mu\nu} = 
-e_\mu^b e^\sigma_a 
\left(\omega_\sigma{}^a{}_c \omega_{\nu}{}^c{}_b - \omega_\nu{}^a{}_c \omega_{\sigma}{}^c{}_b \right)
- \frac{e_\mu^b e_a^\sigma}{4C^2} B_{\sigma,cd}^{a} B_\nu^{bcd}
+ \cdots
\ea
where $R_{\mu\nu}$ is Ricci tensor.
The second term in the left-hand side shows 
this equation describes spacetime with negative cosmological constant.
This seems natural since the near horizon geometry of M-branes is anti-de Sitter spacetime, as is well known. %~[??].  %%% find!!

Let us here note the difference from Einstein gravity.
We first point out that 
the connection $k_\mu^{ab}$ in our formulation is $SO(1,10)$ covariant, 
but in Einstein gravity it is not.
This changes the relation of $k_\mu^{ab}$ and elfbain $e_\mu^a$:
in Einstein gravity it is given as eq.\,(\ref{omega_def}),
while in our formulation it becomes eq.\,(\ref{eom_e}) including the currents $W_\mu$ and  $B_\mu$.
Then we may say that if any effects from branes are negligible, 
our formulation is coincident with Einstein gravity.

As a result, we obtain the geometric equation (\ref{Einstein-like}).
The first term in the right-hand side shows a deviation from Einstein equation,
which can be interpreted as additional gravitational energy.
However, our equation does have general covariance.
The deviation appear only because our connection $k_\mu^{ab}$ has 
different covariance from Einstein gravity.
Moreover, any transformations of the connections and the currents 
cannot absorb this deviation.
This means we derive a nonequivalent equation with Einstein equation.

%and at this time how it should be interpreted is not clear.
%However, if we are setting this aside, we may find 

%%%%%%%%%%%%%%%%%%%%%%%%%%%%%%%%%%%%%%%%%%%%%
\section{M-brane solutions in flat spacetime ($k_\mu^{ab}(x)=0$)}
\label{sec:flat}
%%%%%%%%%%%%%%%%%%%%%%%%%%%%%%%%%%%%%%%%%%%%%

Let us now solve the equations of motion listed in \S\,\ref{sec:EOM}.
However, since they are slightly complicated, 
we first discuss simple cases with the connection field $k_\mu^{ab}=0$. 
This means that in this section we consider only the flat spacetime.

\subsection{M2-brane}

First we discuss a system of only M2-brane, which means 
\ba
W_\mu^{abcde}(x) = W_\mu^{abcdefg,h}(x) = B_\mu^{abcdef}(x) = 0\,.
\ea
Moreover, let us here focus on the currents % (or fields) 
in the worldvolume of this M2-brane, so the metric should be flat:
\ba
k_\mu^{ab}(x) \sim 0\,.
\ea
Note that $\mu$ runs only $0,1,2$ in this subsection.
This means the scale of $k_\mu^{ab}$ is much smaller than that of $C$,
i.e., 11d Planck scale.
In this setting, we find that 
the equations of motion (\ref{eom_M2}), (\ref{eom_B3}) and (\ref{eom_e}) become
\ba
\partial_\mu W_\nu^{ab} - \partial_\nu W_\mu^{ab}
&=& -\frac{2i}{C} B_{[\mu}^{abc} e_{\nu]c} \nt 
\partial_\mu e_\nu^a - \partial_\nu e_\mu^a
&=& \frac{i}{C} B_{[\mu}^{abc} W_{\nu]bc} \nt
\partial_\mu B_\nu^{abc} - \partial_\nu B_\mu^{abc}
&=& \frac{2i}{C} W_{[\mu}^{[ab} e_{\nu]}^{c]}\,,
\ea
and the conservation equations become
\ba
\partial^\mu W_\mu^{ab} = \partial^\mu e_\mu^a = \partial^\mu B_\mu^{abc} = 0 \,.
\ea
Then all these equations are satisfied when 
\ba \label{setM2}
e_\mu^a(x) &=& i \partial_\mu X^a\nt
B_\mu^{abc}(x) &=& 
-i C \partial_\mu \left(\eps^{\nu\rho\sigma} \partial_\nu X^a \partial_\rho X^b \partial_\sigma X^c\right) \nt
% = i C\partial_i (\eps^{jkl} e_j^a e_k^b e_l^c) \,,\nt
W_\mu^{ab}(x) &=& - i \eps_{\mu\nu\rho}\partial^\nu X^a \partial^\rho X^b
 = e_{\mu c} \eps^{\nu\rho\sigma} \partial_\nu X^a \partial_\rho X^b \partial_\sigma X^c.
% = \frac12 \eps_{ijk} e^{ja} e^{kb}\,.
\ea
Here $X^a$ can be regarded as a position of the M2-brane in 11d spacetime.
These degrees of freedom should correspond to the gauge and scalar fields
in the field theory on M2-brane worldvolume:
the gauge field describes the longitudinal directions $a=0,1,2$ for the brane,
while the scalar fields correspond to the transverse directions $a=3,\ldots,10$.
%(we need some comments on $C$ because there are some ambiguity...)
%(so that all the currents $e_i^a, B_i^{abc}, W_i^{ab}$ are Hermitian and dimension 0. 
%The field $X^a$ are also Hermitian.) 
%For more general setups, we need to discuss $k_i^{ab} \neq 0$ cases.

\subsection{M5-brane}

Next we discuss a system of only M5-brane, which means 
\ba
W_\mu^{ab}(x) = B_\mu^{abc}(x) = 0\,.
\ea
Similarly, we focus on the currents in the worldvolume of the M5-brane,
and we consider only the cases where this M5-brane doesn't intersect with any other branes.
%which is expanded in particular directions ().
%that is, neglect all the other M5-branes expanded in different directions.
In this setting, the metric should be flat:
\ba
k_\mu^{ab}(x) \sim 0\,.
\ea
In this subsection, $\mu$ runs only $0,\ldots,5$.
Then we find that the equations of motion (\ref{eom_M5}), (\ref{eom_B6}) and (\ref{eom_e}) become
\ba
\partial_\mu W_\nu^{abcde} - \partial_\nu W_\mu^{abcde}
&=& \frac{2i}{C} B_{[\mu}^{abcdef} e_{\nu]f} \nt   %%% CHECK!!! %%%
\partial_\mu e_\nu^a - \partial_\nu e_\mu^a
&=& -\frac{i}{C} B_{[\mu}^{abcdef} W_{\nu]bcdef} \nt
\partial_\mu B_\nu^{abcdef} - \partial_\nu B_\mu^{abcdef}
&=& -\frac{2i}{C} W_{[\mu}^{[abcde} e_{\nu]}^{f]} \,,
\ea
and the conservation equations become
\ba
\partial^\mu W_\mu^{abcde} = \partial^\mu e_\mu^a 
= \partial^\mu B_\mu^{abcdef} = 0 \,.
\ea
All these equations are satisfied when $e_\mu^a = i \partial_\mu X^a$ and 
\ba \label{setM5}
B_\mu^{abcdef}(x) &=& 
-iC \partial_\mu \left(\eps^{\nu\rho\sigma\kappa\lambda\tau} 
\partial_\nu X^a \partial_\rho X^b \partial_\sigma X^c \partial_\kappa X^d \partial_\lambda X^e \partial_\tau X^f \right) \nt
% &=& iC \partial_i \eps^{jklmnp} e_j^a e_k^b e_l^c e_m^d e_n^e e_p^f\,, \nt
W_\mu^{abcde}(x) &=& 
 i \eps_{\mu\nu\rho\sigma\kappa\lambda} \partial^\nu X^a \partial^\rho X^b \partial^\sigma X^c \partial^\kappa X^d \partial^\lambda X^e \nt
&=& e_{\mu f} \eps^{\nu\rho\sigma\kappa\lambda\tau} \partial_\nu X^a \partial_\rho X^b \partial_\sigma X^c \partial_\kappa X^d \partial_\lambda X^e \partial_\tau X^f .
% &=& \frac{1}{5!} \eps_{ijklmn} e^{ja} e^{kb} e^{lc} e^{md} e^{ne}\,,
\ea
Again, the degrees of freedom $X^a$ should show a position of the M5-brane in 11d spacetime, 
and correspond to the gauge and scalar fields on the M5-brane worldvolume.

%%%%%%%%%%%%%%%%%%%%%%%%%%%%%%%%%%%%%%%%%%%%%%%%%%%%
\section{M-brane solutions in curved spacetime ($k_{\mu}^{ab}(x) \neq 0$)}
\label{sec:curved}
%%%%%%%%%%%%%%%%%%%%%%%%%%%%%%%%%%%%%%%%%%%%%%%%%%%%

In spite of plausible results for the special cases in Sec.\,\ref{sec:flat},
it is at least physically not consistent to write the brane equations as if the spacetime is flat,
since the branes themselves constitute the sources of gravity equations (i.e., equations for $k_\mu^{ab}(x)$). 
In fact, it is not difficult to write the equations and the solutions when $k_\mu^{ab}(x)$ is non-vanishing: 
All we need to do is to replace the derivative $\partial _{\mu }$ by the covariant derivative $D_{\mu }$.

\subsection{Equations for currents}

First, the equations for M2-brane (\ref{eom_M2}) and (\ref{eom_B3}) become
\ba \label{43}
D_{\mu }W_\nu^{ab}(x)-D_{\nu }W_\mu^{ab}(x)
= \frac{i}{C} \left( 
- B_{\mu}^{abc}(x)e_{\nu,c}(x) +\cdots\right)
- (\mu \leftrightarrow \nu ) 
\ea
and
\ba \label{DB3}
D_{\mu }B_\nu^{abc}(x)-D_{\nu }B_\mu^{abc}(x)
&=& \frac{i}{C}\left(
 W_\mu^{[ab}(x) e_\nu^{c]}(x)
%-48 W_{\mu}^{de}(x)W_{\nu}^{abc}{}_{de}(x) 
%- \frac{1}{4}W_{\mu}^{defgh}(x) W_{\nu,defgh}{}^{abc}(x) 
+\cdots \right)
-(\mu \leftrightarrow \nu )\,,
\ea
where $D_{\mu}W_\nu^{ab}(x)$ is defined as
\ba
D_{\mu}W_\nu^{ab}(x)
&=& \partial_{\mu}W_{\nu}^{ab}(x)
+\frac{i}{2C}\left(k_{\mu}{}^a{}_c(x)W_{\nu }^{cb}(x) + k_{\mu}{}^b{}_c(x)W_{\nu }^{ac}(x)\right)
\nt
&=& \partial_\mu W_\nu^{ab}(x) - \frac{i}{C} k_\mu{}^{[a}{}_c(x) W_\nu^{b]c}(x)\,.
\ea

We also have the conservation equations
\ba
D^{\mu }W_\mu^{ab}(x)=D^{\mu }B_\mu^{abc}(x)=0\,.
\ea
Here the caution must be taken when we define the covariant derivative. 
In fact the covariant derivative in the conservation equation is 
\ba
D^\mu = g^{\mu\nu}\partial_\nu - g^{\nu\lambda} \Gamma_{\nu\lambda}^\mu\,.
\ea
This is because $k_{\mu}{}^a{}_c(x) W_{\nu }^{cb}(x)$ or any other similar terms $k_{\mu}{}^a{}_c(x)X_{\nu}^{cbdef\cdots}(x)$ must be antisymmetric for indices $a,b,d,e,f,\ldots$,  which can be derived from the original equation $-i\partial_\mu J_\nu=\left[P_\mu, J_\nu \right]$.
Therefore, the $k_\mu^{ab}$ term does not appear in the conservation equation,
whereas the $\Gamma _{\nu \lambda}^{\mu}$ term does not appear in the antisymmetric equation because of the symmetry
$\Gamma _{\nu \lambda }^{\mu } =\Gamma _{\lambda \nu }^{\mu }$~\cite{Sugawara:1967rw}.

We have the similar definition for $D_{\mu}B_\nu^{abc}(x)$ with each term contracted by $k_{\mu}{}^a{}_b(x)$:
\ba \label{cov_B3}
D_{\mu }B_\nu^{abc}(x)
&=& \partial_{\mu }B_\nu^{abc}(x)+\frac{i}{2C}\left(k_{\mu}{}^a{}_d(x)B_{\nu }^{dbc}(x)
 +k_{\mu}{}^b{}_d(x)B_{\nu }^{adc}(x)+k_{\mu}{}^c{}_d(x)B_{\nu }^{abd}(x)\right) \nt
&=& \partial_\mu B_\nu^{abc}(x) +\frac{3i}{2C} k_\mu{}^{[a}{}_d(x) B_\nu^{bc]d}(x)\,.
\ea
One sees that $B_\mu^{abc}(x)$ plays the role of source term in the equation for $W_\mu^{ab}(x)$, and inversely $W_\mu^{ab}(x)$ plays the role of source term in the equation for $B_\mu^{abc}(x)$. This indicates an aspect of the field-current identity.

Next, in a similar way, 
we have the equations for M5-brane (\ref{eom_M5}) and (\ref{eom_B6}):
\ba \label{eq_DW5}
D_{\mu }W_{\nu }^{abcde}(x)-D_{\nu }W_{\mu }^{abcde}(x)
&=&
\frac{i}{C}\left(
 B_{\mu}^{abcdef}(x)e_{\nu,f}(x)
- B_{\mu}^{[abc}(x) W_{\nu}^{de]}(x)
%+\frac1{24} B_\mu^{fgh}(x) W_\nu^{abcde}{}_{fg,h}(x)
+\cdots \right)
\nt && \quad
-(\mu \leftrightarrow \nu)
\ea
and 
\ba \label{DB6}
D_{\mu }B_{\nu }^{{abcdef}}(x)-D_{\nu }B_{\mu }^{{abcdef}}(x)
&=&
\frac{i}{C}\left(
- W_{\mu}^{[abcde}(x)e_{\nu}^{f]}(x)
-40 B_\mu^{[abc}(x) B_\nu^{def]}(x)
%+\frac{7}{10} W_\mu^{abcgh}(x) W_\nu^{def}{}_{gh}(x)
%+15 W_\mu^{gh} W_\nu^{abc}{}_{gh}{}^{de,f} 
+\cdots \right)
\nt && \quad
-(\mu \leftrightarrow \nu)\,.
\ea

We also have the conservation equations just as in the case of M2-brane,
\ba
D^{\mu }W _\mu^{abcde}(x)=D^{\mu }B_\mu^{abcdef}(x)=0\,.
\ea
Similar caution must be taken when we define the covariant derivative in the conservation equations, as in the case of M2-brane.
The definition of the covariant derivatives for antisymmetric currents $W_\mu^{abcde}(x)$ and $B_\mu^{abcdef}(x)$ are the same as in the case of M2-brane, 
so we don't write it down 
% and will not be written down
explicitly here.

If we consider the situations where M2- or M5-brane exists alone, similarly to Sec.\,\ref{sec:flat},
%we can show %%%%%%% (?)
all the above equations are satisfied by the following ansatz:
\ba \label{ansatz}
W _{\mu }^{{ab}}(x)
&=& e_{{\mu c}}(x)\epsilon ^{\nu \rho \sigma }D_{\nu }X^aD_{\rho }X^bD_{\sigma }X^c \nt
W _{\mu }^{{abcde}}(x)
&=& e_{{\mu f}}(x)\epsilon ^{\nu \rho \sigma \kappa \lambda \tau }D_{\nu }X^aD_{\rho}X^bD_{\sigma }X^cD_{\kappa }X^dD_{\lambda }X^eD_{\tau }X^f \nt
B_{\mu }^{{abc}}(x)
&=& -iCD_{\mu }\left(\epsilon ^{\nu \rho \sigma }D_{\nu }X^aD_{\rho }X^bD_{\sigma }X^c\right) \nt
B_{\mu }^{{abcdef}}(x)
&=& -iCD_{\mu }\left(\epsilon ^{\nu \rho \sigma \kappa \lambda \tau }D_{\nu }X^aD_{\rho }X^bD_{\sigma }X^cD_{\kappa}X^dD_{\lambda }X^eD_{\tau }X^f\right),
\ea
together with
\ba \label{20}
e_{\mu }^a(x) = iD_{\mu }X^a
= i\left(\delta _b^a\partial _{\mu }+\omega_\mu{}^a{}_b \right)X^b.
\ea
Here the anti-hermitian nature of these currents are taken into account by assignment of an appropriate factors. 
We can easily find this ansatz is reduced to eqs.\,(\ref{setM2}) and (\ref{setM5}) in the limit of  the flat spacetime ($D_\mu \to \partial_\mu$).

\subsection{Solutions of M-branes}

Now we solve the equations for the currents in more general situations.
The ansatz (\ref{20}) for $e_{\mu}^{a} (x)$ must be justified by satisfying the geometric equations (i.e., equations for $k_{\mu}^{ab}(x)$ and $e_{\mu}^{a}(x)$), 
but we will discuss it in the next section. 
Here we simply check one of the geometric equations:
\ba
D^{\mu} e_{\mu}^{a} (x) = 0
\ea
with 
$D^{\mu} = g^{\mu \nu} \partial_{\nu} - g^{\nu \lambda} \Gamma^\mu_{\nu\lambda}$
and
$\Gamma^\mu_{\nu\lambda} = \frac{1}{2} e_{a}^{\mu} (\partial_{\nu} e_{\lambda}^{a} + \partial_{\lambda} e_{\nu}^{a})$.
We note that the $k_{\mu}^{ab}$ contribution to the covariant term does not appear because of the antisymmetric nature of $k_{\mu}{}^a{}_b e_{\nu}^{b}$, which is imposed as the supplementary condition as discussed before.
Then we get
\ba \label{A}
\left( g^{\mu \nu} \partial_{\nu} - g^{\nu \lambda} ie_{a}^{\mu} \partial_{\nu} D_{\lambda} X^{a} \right) D_{\mu} X^{c} = 0\,.
\ea
By putting
$g^{\nu \lambda} \partial_{\nu} D_{\lambda} X^{a} =: Y^{a}$,
this equation % eq.\,(\ref{A}) 
becomes
\ba
    Y^{c} - ie_{a}^{\mu} Y^{a} D_{\mu} X^{c} = 0\,.
\label{B}
\ea
Since we have eq.\,(\ref{20}), or equivalently 
\ba
    ie_{a}^{\mu} D_{\mu} X^{c} = \delta_{a}^{c}
\ea
in this approximation, we find that eq.\,(\ref{A}) is trivially satisfied.

We now return to the main theme of this subsection:
the M2- and M5-brane solutions.
The ansatz (\ref{ansatz}) for the brane currents and the brane charge currents 
should be generalized by adding terms corresponding to the non-linear realization.
%The M2-brane solution is characterized by
%\ba
%W_{\mu}^{ab} (x) &=& \frac{1}{2} e_{\mu c} (x) \epsilon^{\nu \rho \sigma} D_{\nu} X^{a} D_{\rho} X^{b} D_{\sigma} X^{c} 
%+ \alpha D_{\mu} \left[ \epsilon^{\nu \rho} D_{\nu} X^{a} D_{\rho} X^{b} \right] 
%+ \cdots \nt
%W_{\mu}^{abcde} (x) 
%&=& \alpha e_{\mu f} (x) \epsilon^{\nu \rho \sigma \kappa \lambda \tau} D_{\nu} X^{a} D_{\rho} X^{b} D_{\sigma} X^{c} D_{\kappa} X^{d} D_{\kappa} X^{d} D_{\lambda} X^{e} D_{\tau} X^{f} + \cdots \nt
%\ea
Especially, the M5-brane solution should be characterized by
\ba
W_{\mu}^{abcde} (x) 
&=& \alpha e_{\mu f} (x) \epsilon^{\nu \rho \sigma \kappa \lambda \tau} D_{\nu} X^{a} D_{\rho} X^{b} D_{\sigma} X^{c} D_{\kappa} X^{d} D_{\kappa} X^{d} D_{\lambda} X^{e} D_{\tau} X^{f} \nt
&&
+\, \beta_\tau D_{\mu} \left[\epsilon^{\nu \rho \sigma \kappa \lambda \tau } D_{\nu} X^{a} D_{\rho} X^{b} D_{\sigma} X^{c} D_{\kappa} X^{d} D_{\lambda} X^{e} \right] + \cdots,
\ea
where $e_{\mu c} (x)$ is given by $D_{\mu} X_{c}$. 
Here we add extra terms % to %$W_{\mu}^{ab} (x)$ and to $W_{\mu}^{abcde} (x)$ 
with constant parameters $\alpha$ and $\beta_\mu$,
so that we can satisfy the equations for these variables in more general cases.
The constant vector $\beta_\mu$ should be closely related to the auxiliary field on M5-branes in the famous PST formulation~\cite{Pasti:1997gx},
and we will clarify this point in a future work.
%These terms correspond to the non-linear realization of these terms.

%We also have the expressions of the brane charge currents as
%\ba
%B_{\mu}^{abc} (x) = -iC D_{\mu} \left( \epsilon^{\nu \rho \sigma} D_{\nu} X^{a} D_{\rho} %X^{b} D_{\sigma} X^{c} \right) + \cdots\,,
%\ea
%and
%\ba \label{B6_curved}
%B_{\mu}^{abcdef} (x) = - iCD_{\mu} \left( \epsilon^{\nu \rho \sigma \kappa \lambda \tau}  %D_{\nu} X^{a} D_{\rho} X^{b} D_{\sigma} X^{c} D_{\kappa} X^{d} D_{\lambda} X^{e} D_{\tau} %X^{f} \right) + \cdots\,.
%\ea

If we can ignore such non-linear realization terms, 
using the ansatz (\ref{ansatz}), 
the equation for M2-brane current (\ref{43}) becomes
%we have 
\ba
&&D_{\mu} W_{\lambda}^{ab} (x) - D_{\lambda} W_{\mu}^{ab} (x) 
\nt
&&= \left[ 
D_{\mu} e_{\lambda c} (x) %- D_{\lambda} e_{\mu c} (x) \right) 
\epsilon^{\nu \rho \sigma} D_{\nu} X^{a} D_{\rho} X^{b} D_{\sigma} X^{c} 
% +\, e_{\lambda c} (x) D_{\mu} \left( \epsilon^{\nu \rho \sigma} D_{\nu} X^{a} D_{\rho} X^{b}  D_{\sigma} X^{c} \right)
 - e_{\mu c} (x) D_{\lambda} \left( \epsilon^{\nu \rho \sigma} D_{\nu} X^{a} D_{\rho} X^{b} D_{\sigma} X^{c} \right) + \cdots \right]
\nt && \quad
- (\mu \leftrightarrow \lambda)
\nt
&&= \frac{i}{C} \left( e_{\mu c} (x) B_{\lambda}^{abc} (x) + \cdots \right)
- (\mu \leftrightarrow \lambda).
\label{4.19}
\ea
The first term in the intermediate step of eq.\,(\ref{4.19}) vanishes,
because one of the geometric equations for $e_{\mu c} (x)$ which we will discuss in the next section is
\ba \label{eom_e_curved}
D_\mu e_{\lambda c} (x) - D_{\lambda} e_{\mu c} (x) = 0\,.
\ea
Therefore, the equation for $W_{\mu}^{ab} (x)$ is satisfied. 
Compared with the previous section, 
the inclusion of $k_{\mu}^{ab}$ terms can be done only by replacing all the derivatives 
$\partial_{\mu}$ by $D_{\mu}$ in above equations. No other change is needed.

Next, we discuss the equations for M2-brane charge current (\ref{DB3}):
\ba
D_{\mu} B_{\nu}^{abc} (x) - D_{\nu} B_{\mu}^{abc} (x) 
= \frac{i}{C} \left( W_{\mu}^{[ab} (x) e_{\nu}^{c]} (x) + \cdots\right)
- (\mu \leftrightarrow \nu )
\label{4.20}
\ea
where the covariant deriative $D_{\mu} B_{\nu}^{abc} (x)$ has been defined in eq.\,(\ref{cov_B3}).
%= \partial_{\mu} B_{\nu}^{abc} (x) + \frac{i}{2C} k_{\mu}{}^{[c}{}_d B^{ab] d}_{\nu} 
%= \partial_{\mu} B_{\nu}^{abc} (x) + 2 \omega_{\mu}{}^{[c}{}_d B^{ab] d}_{\nu}.
%\label{4.21}
We define here 
\ba
B^{abc} := \epsilon^{\nu \rho \sigma} D_{\nu} X^{a} D_{\rho} X^{b} D_{\sigma} X^{c},
\ea
then we obtain 
\ba
B_{\mu}^{abc} (x) &=& - i CD_{\mu} \left( \epsilon^{\nu \rho \sigma} D_{\nu} X^{a} D_{\rho} X^{b} D_{\sigma} X^{c} \right)
= - i CD_{\mu} B^{abc} \nt
W_{\mu}^{ab} (x) 
&=&  e_{\mu c} (x) \epsilon^{\nu \rho \sigma} D_{\nu} X^{a} D_{\rho} X^{b} D_{\sigma} X^{c}
=  e_{\mu c} (x) B^{abc}.
%\label{4.23}
\ea
Using these expressions, eq.\,(\ref{4.20}) can be written as
\ba
( D_{\mu} D_{\nu} - D_{\nu} D_{\mu} ) B^{abc} 
= -\frac{1}{C^2} \left( e_{\mu d} e_{\nu}^{[c} B^{ab]d} +\cdots\right)
- ( \mu \leftrightarrow \nu) .
%\label{4.24}
\ea
%or
%\ba
%\left[ 
%( \partial_{\mu} \delta_{e}^{[c} + 2 \omega_{\mu}{}^{[c}{}_e ) (\partial_{\nu} \delta_{d}^{e} + 2 \omega_{\nu}{}^e{}_d) - ( \mu \leftrightarrow \nu) \right] B^{ab]d} 
%= -\frac{1}{2C^{2}} \left( e_{\mu d}  e_{\nu}^{[c} B^{ab] d} - (\mu \leftrightarrow \nu) \right)\,.
%\ea
This equation can be satisfied, if we use the ansatz (\ref{20}) and we have
%\ba
%( \partial_{\mu} \delta_{e}^{c} + 2 \omega_{\mu}{}^c{}_e ) (\partial_{\nu} \delta_{d}^{e} + 2 \omega_{\nu}{}^e{}_d ) - (\mu \leftrightarrow \nu) 
%= -\frac{1}{2C^{2}} \left(  e_{\mu d} (x) e_{\nu}^{c} - (\mu \leftrightarrow \nu) \right).
%\ea
%That is,
\ba
 ( \partial_{\mu} \omega_{\nu}{}^c{}_d + 2 \omega_{\mu}{}^c{}_e \omega_{\nu}{}^e{}_d ) - (\mu \leftrightarrow \nu) 
= \frac{1}{2C^{2}} (e_{\mu d} e_{\nu}^{c} - e_{\nu d} e_{\mu}^{c} ) + \cdots. 
\ea
By using the relation 
$\omega_{\mu}^{ab} (x) = \frac{i}{2C} k_{\mu}^{ab} (x)$,
this equation is rewritten as 
\ba
\left( \partial_{\mu} k_{\nu}{}^c{}_d + \frac{i}{C} k_{\mu}{}^c{}_e k_{\nu}{}^e{}_d\right) 
- (\mu \leftrightarrow \nu) 
&=& -\frac{i}{C} ( e_{\mu d} e_\nu^{c} - e_{\nu d}  e_{\mu}^{c} ) + \cdots.
\ea
This should be compared with one of the geometric equations, 
namely, the equation for $k_{\mu}{}^a{}_b$, which we will discuss in the next section,
\ba \label{eom_k_curved}
&&  %%% coeff: OK %%%
D^{ab, cd}_{\mu}  k_{\nu, cd} (x) - D^{ab, cd}_{\nu} k_{\mu, cd} (x) 
\nt
&&= \frac{i}{C} \left( e^{a}_{\mu} (x) e_{\nu}^{b} (x) - e^{b}_{\mu} (x) e_{\nu}^{a} (x) \right) 
- \frac{i}{C} \left( W_{\mu}^{[a}{}_d (x) W_{\nu}^{b]d} (x) - W_{\nu}^{[a}{}_d (x) W_{\mu}^{b]d} (x) \right) \nt
&& \quad
+\, \frac{i}{4!C} \left(W_{\mu}^{[a}{}_{defg} (x) W_{\nu}^{b]defg} (x) - W_{\nu}^{[a}{}_{defg} (x) W_{\mu}^{b]defg} (x)\right) + \cdots
%\label{4.26}
\ea
where we define
\ba
D^{ab,cd}_\mu := \eta^{ac}\eta^{bd}\partial_\mu + \frac{i}{2C}(\eta^{ad}k_\mu^{bc}+\eta^{bc}k_\mu^{ad}).
\ea
This equation (\ref{eom_k_curved}) shows 
that the equation for M2-brane charge current $B_{\nu }^{abc} (x)$ can be satisfied 
if the equation for $k^{ab}_{\mu} (x)$ is satisfied,
when the contribution of the branes to the latter equation can be ignored. 

The next task is to prove the M5-brane equations are satisfied by our ansatz.
The equation for M5-brane current (\ref{eq_DW5}) is 
\ba  %%% coeff %%%
&&
D_{\mu} W_{\nu}^{abcde} (x) - D_{\nu} W_{\mu}^{abcde} (x) 
\nt
&&= 
\frac{i}{C} \left(
 B_{\mu}^{abcdef}(x) e_{\nu f} (x)
- B^{[abc}_{\mu} (x) W_{\nu}^{de]} (x) 
%+ \frac{1}{24} B_{\mu}^{ghl} (x) W_{\nu}^{abcde}{}_{gh,l} (x) 
+ \cdots \right)  
- ( \mu \leftrightarrow \nu)\,.
\ea
Inserting our ansatz (\ref{ansatz}), we get
\ba
&&
D_{\mu} W_{\xi}^{abcde} (x) - D_{\xi} W_{\mu}^{abcde} (x) 
\nt &&
= e_{\xi f} (x) D_{\mu} \left( \epsilon^{\nu \rho \sigma \kappa \lambda \tau} D_{\nu} X^{a} D_{\rho} X^{b} D_{\sigma} X^{c} D_{\kappa} X^{d} D_{\lambda} X^{e} D_{\tau} X^{f} \right)
- (\mu\leftrightarrow \xi) \nt
%-\, e_{\mu f} (x) \epsilon^{\nu \rho \sigma \kappa \lambda \tau} D_{\xi} { D_{\nu} X^{a} D_{\rho} X^{b} D_{\sigma} X^{c} D_{\kappa} X^{d} D_{\lambda} X^{e} D_{\tau} X^{f} } \nt
&&= \frac{i}{C} B_{\mu}^{abcdef}(x) e_{\xi f} (x) - (\mu \leftrightarrow \xi),
\label{4.27}
\ea
where the use is made of eq.\,(\ref{eom_e_curved})   % make use of...
%\ba
%D_{\mu} e_{\lambda c} (x) - D_{\lambda} e_{\mu c} (x) = 0
%\ea
and the expression
\ba \label{B6_curved}
B_{\mu}^{abcdef} (x) = - iCD_{\mu} \left( \epsilon^{\nu \rho \sigma \kappa \lambda \tau}  D_{\nu} X^{a} D_{\rho} X^{b} D_{\sigma} X^{c} D_{\kappa} X^{d} D_{\lambda} X^{e} D_{\tau} X^{f} \right) + \cdots
\ea
as in the case of M2-brane.
Then eq.\,(\ref{4.27}) can be rewritten as %modified to:
\ba   %%% coeff %%%
\label{DW5}
&&
D_{\mu} W_{\xi}^{abcde} (x) - D_{\xi} W_{\mu}^{abcde} (x) 
\nt &&
= e_{\xi f} (x) \epsilon^{\nu \rho \sigma \kappa \lambda \tau}  
\left[ D_{\nu} X^{a} D_{\rho} X^{b} D_{\sigma} {X^{c}} D_{\mu} \left( D_{\kappa} X^{d} D_{\lambda} X^{e} D_{\tau} X^{f} \right) 
\right. \nt && \left. \qquad
+ D_{\mu} \left( D_{\nu} X^{a} D_{\rho} X^{b} D_{\sigma} X^{c} \right) D_{\kappa} X^{d} D_{\lambda} X^{e} D_{\tau} X^{f} \right] - (\mu \leftrightarrow \xi)  
\nt
&&= \frac{i}{C} { e_{\xi f} (x) \left(B^{[abc} B_{\mu}^{def]} + B_{\mu}^{[abc} B^{def]} \right) 
- ( \mu \leftrightarrow \xi) } 
\nt
&&= \frac{i}{C} { \left( B_{\mu}^{[abc}(x) W_{\xi}^{de]}(x) + e_{\xi f} (x) B^{[abc} (x) B_{\mu}^{def]} (x) \right)
- (\mu \leftrightarrow \xi) }.
%\label{4.28}
\ea
Note that we defined 
$B^{abc} := \epsilon^{\nu \rho \sigma} D_{\nu} X^{a} D_{\rho} X^{b} D_{\sigma} X^{c}$ 
and so this is a hermitian operator.

Let us now comment on the non-linear realization terms:
Our ansatz (\ref{ansatz}) can be generalized by adding suitable terms.
The above equation (\ref{DW5}) suggests that we may have to add an extra term as follows:
\ba \label{eq_W5}
W_{\mu}^{abcde} (x) 
&=& \alpha e_{\mu f} (x)  \epsilon^{\nu \rho \sigma \kappa \lambda \tau} D_{\nu} X^{a} D_{\rho} X^{b} D_{\sigma} X^{c} D_{\kappa} X^{d} D_{\lambda} X^{e} D_{\tau} X^{f} 
\nt &&
-\, \gamma B^{[abc} (x) W_{\mu}^{de]} (x)
\ea
where $\alpha$ and $\gamma$ are constants.
The second term on the right-hand side shows the existence of M2-branes completely 
wrapped by the M5-brane.
The first term gives contributions to eq.\,(\ref{DW5})
\ba
\alpha\frac{i}{C} \left[ \left( B_{\mu}^{[abc}(x) W_{\xi}^{de]}(x) + e_{\xi f} (x) B^{[abc}(x) B_{\mu}^{def]}(x) \right) - (\mu \leftrightarrow \xi) \right],
\ea
and the second term gives 
\ba
%&&
%D_{\mu} W_{\xi}^{abcde} (x)  - D_{\xi} W_{\mu}^{abcde} (x) 
%\nt
&&
- \gamma \left[\left( \frac{i}{C} B_{\mu}^{[abc} (x) W_{\xi}^{de]} (x) + B^{[abc} (x) D_{\mu} W_{\xi}^{de]} (x) \right) - (\mu \leftrightarrow \xi) \right] \nt
&&
= - \gamma \frac{i}{C} \left[ \left( 
B_{\mu}^{[abc} (x) W_{\xi}^{de]} (x) - B^{[abc}(x) B_{\mu}^{de]f} (x) e_{\xi f} (x) \right)
- (\mu \leftrightarrow \xi) \right].
\ea
Therefore, putting all the expressions together, we have the equations
\ba
\alpha\frac{i}{C} - \gamma \frac{i}{C} = -\frac{i}{C}\,, \quad
\alpha\frac{i}{C} + \gamma \frac{i}{C}  = 0\,,
\ea
and easily get the solution $\alpha = -\frac12$, $\gamma = \frac12$.
%with $\alpha =: \frac{1}{C} \tilde{\alpha}$, we get
%\ba
 %   2 \beta - \tilde{\alpha} = 2, \ \ - \beta + \tilde{\alpha} = 0, \ \ \therefore \beta = \tilde{\alpha} = 2.
%\ea
In a similar manner, we may further modify the expression of $W_{\mu}^{abcde}$,
so that it satisfies the equation for non-linearly realized parts of $W_{\mu}^{abcde}$,
by adding the following terms:
%\ba
 %   W_{\mu}^{abcde} (x) = - iD_{\mu} [ W_{\mu}^{abcde} (x) ] - \frac{1}{4C} {\epsilon^{ghl[abcd}}{}_{ijkm} B_{\mu ghl} (x) [ W^{abcde} (x) ] \dots
%\ea
\ba
D_{\mu} W_{\xi}^{abcde} (x) - D_{\xi} W_{\mu}^{abcde} (x) 
= \cdots + 
\frac{i}{4!C} {\epsilon^{fgh[abcd}}{}_{ijkl} B_{\mu,fgh} (x) W_{\xi}^{e]ijkl} (x) - (\mu \leftrightarrow \xi). \nt
\ea
This term should be interpreted as contributions from the intersecting M5-branes 
sharing one spatial direction (denoted as the ``$e$'' direction).
Such a brane configuration is not allowed by the intersection rule~\cite{Tseytlin:1996bh}:
intersecting M5-branes share three spatial directions.
However, as we saw in Sec.\,\ref{sec:E11}, our formulation derives Einstein-like equation
with some deviation, which makes us possible to obtain such an exotic configuration.
We will revisit this point in the next section.
%The better interpretation may be that this is the equation for the intersecting M5-branes along the ``$e$ direction.''

%%%%%%%%%%%%%%%%%%%%%%%%%%%%%%%%%%%%%%
\section{Geometric equations and discussion of solutions}
\label{sec:geometric}
%%%%%%%%%%%%%%%%%%%%%%%%%%%%%%%%%%%%%%%%

We expect that the spacetime will not be flat when we have M2- and/or M5-branes except for very special cases. 
Therefore, it is very important to discuss the geometric equations when the branes exist. In our scheme, the geometric equations or the gravity equations are descried by two currents: elfbein current $e_{\mu}^{ab} (x)$ and spin connection current $\omega_{\mu}^{ab} (x)$. 
The latter is related to the generator of $E_{11}$ algebra by
\ba
\omega_{\mu}^{ab} (x) = \frac{i}{2C} k_{\mu}^{ab} (x)\,.
\ea
In fact, $k_{\mu }^{ab} (x)$ corresponds to the $SO(10, 1)$ subgroup of $E_{11}$.

Using the generic equation of motion 
$-i \partial _{\mu } J_{\nu } = [ P_{\mu },  J_{\nu } ]$,
the $E_{11}\otimes \ell_1$ commutation relations
and our ansatz (\ref{ansatz}),
we can easily write down the geometric equations as follows:
\ba
D_{\mu} e_{\lambda c} (x) - D_{\lambda} e_{\mu c} (x) = 0
\label{5.1}
\ea
and
\ba
&&
 D^{ab,cd}_{\mu} k_{\nu, cd} (x) - D^{ab,cd}_{\nu} k_{\mu, cd} (x) 
\nt
&&= \frac{i}{C} \left( e^{a}_{\mu} (x) e_{\nu}^{b} (x) - e^{b}_{\mu} (x) e_{\nu}^{a} (x) \right) 
- \frac{i}{C} \left( W_{\mu}^{[a}{}_d (x) W_{\nu}^{b]d} (x) - W_{\nu}^{[a}{}_d (x) W_{\mu}^{b]d} (x) \right) \nt
&& \quad
+\, \frac{i}{4!C} \left(W_{\mu}^{[a}{}_{defg} (x) W_{\nu}^{b]defg} (x) - W_{\nu}^{[a}{}_{defg} (x) W_{\mu}^{b]defg} (x)\right) + \cdots.
\label{5.2}
\ea
Here we are using anti-hermitian $e^a_{\mu} (x)$, $k_\mu^{ab}(x)$, $W_{\mu}^{ab} (x)$ and $W_{\mu}^{abcde} (x)$ in this equation.
The signs on the left-hand side of this equation will be changed 
when we use the hermitian variables by dividing each variable by $i$.
We can prove that the contribution of the brane charge currents $B_\mu^{abc}(x)$ and $B_\mu^{abcdef}(x)$ to eq.\,(\ref{5.2}) vanishes, if we use our ansatz (\ref{ansatz}),
since these currents are of the form of total derivative $D_\mu(\cdots)$.
%  . (??) under our ansatz $D_\mu$ proportional to ...
In addition, we also have the conservation equations:
\ba
   D^{\mu } e^{a}_{\mu } (x) = 0\,, \quad
   D^{\mu } k^{ab}_{\mu }(x) = 0\,.
\ea

From now on, we use the hermitian variables for $e^{a}_{\mu} (x)$, $\omega_\mu^{ab}(x)$, $W_{\mu}^{ab} (x)$ and $W_{\mu}^{abcde} (x)$ to discuss the gravity theory.
Then, as in usual gravity theory, eq.\,(\ref{5.1}) can be solved easily to provide the spin connection in terms of the elfbein:
\ba
\Omega ^{bca} = e^{\mu b} e^{\lambda c} \left( \partial _{\mu } e_{\lambda }^{a} - \partial _{\lambda } e_{\mu }^{a} \right) ,
%\label{5.3}
\quad
\Omega ^{bca} + \Omega^{abc} - \Omega^{cab} = -2 e^{\mu b} \omega_{\mu}^{ac}\,.
\label{5.4}
\ea
This equation %Eq.\,(\ref{5.4}) 
gives
\ba
k_{\mu}^{ac} 
= - 2i C \omega_\mu^{ac}
= i C e_{\mu b} \left( \Omega^{bca} + \Omega^{abc} - \Omega^{cab} \right).
\ea

We next discuss eq.\,(\ref{5.2}) with our ansatz (\ref{ansatz}):
\ba
W_{\mu}^{ab} (x) 
&=& e_{\mu c} (x) \epsilon^{\nu \rho \sigma} D_{\nu} X^{a} D_{\rho} X^{b} D_{\sigma} X^{c} 
= e_{\mu c} (x) \epsilon^{abc} \det \left[ D_{\mu} X^{a} \right]
\nt
&=& e_{\mu c} (x) \epsilon^{abc} \det \left[ e_{\mu}^{a} \right]
=: e_{\mu c} (x) \epsilon^{abc} V
%\label{5.6}
\ea
and
\ba
W_{\mu}^{abcde} (x) 
&=& e_{\mu f} (x) \epsilon^{\nu \rho \sigma \kappa \lambda \tau}D_{\nu} X^{a} D_{\rho} X^{b} D_{\sigma} X^{c} D_{\kappa} X^{d} D_{\lambda} X^{e} D_{\tau} X^{f} \nt
&=& e_{\mu f} (x) \epsilon^{abcdef} \det \left[ D_{\mu} X^{a} \right] 
= e_{\mu f} (x) \epsilon^{abcdef} \det \left[ e_{\mu}^{a} \right] \nt
&=:& e_{\mu f} (x) \epsilon^{abcdef} H \,.
%\label{5.7}
\ea
Then we can rewrite eq.\,(\ref{5.2}) as
\ba
&&
\partial_{\mu} \omega_\nu{}^{a}{}_{b} (x) - \partial_{\nu} \omega_\mu{}^{a}{}_{b} (x)
+ 2 \omega_\mu{}^{a}{}_{d} \omega_\nu{}^{d}{}_{b} (x) 
- 2 \omega_\nu{}^{a}{}_{d} \omega_\mu{}^{d}{}_{b} (x) 
\nt &&
= \frac{1}{2C^{2}} \left( e_{\mu}^{a} e_{\nu b} - e_{\nu}^{a} e_{\mu b} \right) 
\left( 1 - \frac{1}{4} V^{2} \right) 
\label{5.8}
\ea
for the M2-brane case, and
\ba
&&
\partial_{\mu} \omega_\nu{}^{a}{}_{b} (x) - \partial_{\nu} \omega_\mu{}^{a}{}_{b} (x)
+ 2 \omega_\mu{}^{a}{}_{d} \omega_\nu{}^{d}{}_{b} (x) 
- 2 \omega_\nu{}^{a}{}_{d} \omega_\mu{}^{d}{}_{b} (x) 
\nt &&
= \frac{1}{2C^{2}} \left( e_{\mu}^{a} e_{\nu b} - e_{\nu}^{a} e_{\mu b} \right) 
\left( 1 - \frac{1}{4} H^{2} \right) .
\label{5.9}
\ea
for the M5-brane case.

We can consider the case where we have both M2- and M5-branes with the M2-brane completely wrapped inside the M5-brane in the following way. 
Suppose that M5-brane is expanded along the $1, 2, 3, 4, 5$ directions 
and the M2-brane is along the $4, 5$ directions completely wrapped inside the M5-brane.
Then the equation for the $0, 4, 5$ directions reads %to, as, ...
\ba
&&
\partial_{\mu} \omega_{\nu}{}^a{}_b (x) - \partial_{\nu} \omega_\mu{}^a{}_b (x) 
+ 2 \omega_\mu{}^{a}{}_{d} \omega_\nu{}^{d}{}_{b} (x) 
- 2 \omega_\nu{}^{a}{}_{d} \omega_\mu{}^{d}{}_{b} (x) \nt
\nt
&& =  \frac{1}{2C^{2}} \left( e_{\mu}^{a} e_{\nu b} - e_{\nu}^{a} e_{\mu b} \right) 
\left[ 1 - \frac{1}{4} ( V^{2} + H^{2} ) \right],
%\label(5.10)
\ea
while the equation for the $1, 2, 3$ directions becomes
\ba
&&
\partial_{\mu} \omega_{\nu}{}^a{}_b (x) - \partial_{\nu} \omega_\mu{}^a{}_b (x) 
+ 2 \omega_\mu{}^{a}{}_{d} \omega_\nu{}^{d}{}_{b} (x) 
- 2 \omega_\nu{}^{a}{}_{d} \omega_\mu{}^{d}{}_{b} (x) \nt
&&=  \frac{1}{2C^{2}} \left( e_{\mu}^{a} e_{\nu b} - e_{\nu}^{a} e_{\mu b} \right)
\left( 1 - \frac{1}{4} H^{2} \right).
\label{5.11}
\ea
It is easy to check that eqs.\,(\ref{5.8}) and (\ref{5.9}) have static solutions,
indicating that we can have a flat spacetime when M2- or M5-brane exists alone. 
We can also check that, %in the situation 
when both M2- and M5-branes exist in the configuration mentioned above, 
namely, in the configuration where M2-brane is wrapped by M5-brane, 
there is also a static solution.
%where the spacetime is flat. 

One may wonder if it is consistent with the intersection rule %theorem 
where it is proven that there is a gravity solution only when M2- and M5-branes share one direction~\cite{Tseytlin:1996bh}.
In fact, the reason we have the solution in the configuration of M5-brane completely wrapping M2-brane is because we are not solving the usual Einstein equation but its modified version.

Let us explain this point a little more in detail.
Our gravity equation, for example, eq.\,(\ref{5.11}) is the equation in 4d spacetime. 
Therefore, it is important to compare it with the Einstein equation. 
By using the usual expression for the curvature in terms of spin connection 
$\omega_\mu{}^a{}_b (x)$, eq.\,(\ref{Riemann}), we can rewrite eq.\,(\ref{5.11}) as the following equation:
\ba
R_{\sigma \nu} + e_{a}^{\mu} e_{\sigma}^{b} 
\left( \omega_{\mu}{}^a{}_d \omega_{\nu}{}^d{}_b (x) - \omega_{\nu}{}^a{}_d \omega_\mu{}^d{}_a (x) \right) 
= \frac{5}{C^{2}} g_{\sigma \nu} \left[ 1 - \frac{1}{4} ( V^{2} + H^{2} ) \right].
\ea
We note that spin connection cannot be totally absorbed into the curvature tensor $R_{\mu\nu}$ due to the fact that spin connection in our case corresponds to the $SO(10, 1)$ generators.
It gives %shows that we have 
this extra contribution --- we may say the contribution of graviton itself --- to the spacetime curvature.

The static and also the non-static solutions of eq.\,(\ref{5.11}) will be discussed in our subsequent papers.

%%%%%%%%%%%%%%%%%%%%%%%%%
\section{Concluding remarks}
%%%%%%%%%%%%%%%%%%%%%%%%%

In this paper, we described how M2- and M5-branes can be incorporated into the $E_{11}$ current algebra formulation of M-theory. 
The role of $E_{11}$ algebra in M-theory was first pointed out by P. West and his collaborators~\cite{West:2003fc},
and what we are doing in this paper %here
is to quantize their theory using the current algebra technique~\cite{Sugawara:1967rw} rather than %to refer to 
the ordinary canonical or path integral formulation. 
The M2- and M5-branes are connected to the fact that
we have the second and fifth rank antisymmetric tensors in the $E_{11}$ representation. 

Of course, any representation of $E_{11}$ Kac-Moody algebra is infinite dimensional,
and here we are picking up a few of the simplest components of a representation.
Therefore, if we maintain our theory to be {\em covariant} under $E_{11}$ algebra
with infinite dimensional representations,
our M2- or M5-branes should correspond to infinite multiple branes.
If there is a subgroup of $E_{11}$ algebra (which we call $X(N)$ here) that gives finite degrees of freedom $N^{\frac32}$ to the M2-branes and $N^3$ to the M5-branes,
we can restrict our theory to describe a finite number of branes
by making our theory $X(N)$ covariant.
However, at this time we don't know whether there is such a subgroup. 
%This must be an important future work.
For the same reason, % In fact, 
the spacetime must be also infinite dimensional in $E_{11}$ covariant theory, 
just as supersymmetric theory has the superspace coordinates 
in addition to the regular bosonic spacetime coordinates. 
In this paper, we simply picked up only the $(10+1)$-dimensional part.

% Our M2- or M5-branes corresponds to infinite multiple branes and thus the symmetry 
% is $E_{11}$ with infinite dimensional representations.
% In fact, our spacetime must be also infinite dimensional 
% but 

Thus we study only a simple part of the $E_{11}$ current algebra formulation in this paper, 
but we can successfully show that 
the equations of motion for these variables seem to be satisfied by certain expressions written in terms of the brane coordinates. 
Moreover, the geometry or the gravity can be correctly described by the two currents: elfbein and spin connection, at least at the level of our application.
The full $E_{11}$ covariant theory must include all the components of the infinite dimensional representation. 

There are several directions along which our future work must be done:

\vspace{7pt}
\noindent
(1) Clarify the relation between the Kac-Moody algebra and the diffeomorphism for which there has been already some works done~\cite{Frappat:1989gn}.

\noindent
(2) Work on F-theory as the $E_{12}$ current algebra theory. 
We are not aware of any work toward this direction.

\noindent
(3) Solve eqs.\,(\ref{5.8}) and (\ref{5.9}) to find a configuration 
which can describe our real 4d spacetime (in particular, exponentially expanding universe).
Our static solution in Sec.\,\ref{sec:geometric} shows 
that we may need some extra currents in addition to the brane currents $W_\mu$ 
and the brane charge currents $B_\mu$ to make our $(3+1)$-dimensional
spacetime (i.e., M5-brane wrapping M2-brane) time-dependent.

\begin{comment}
Here we may make a few comments regarding (3). We checked that ordinary homogeneous and isotropic universe may not be the solution to these
equations. That means that the simple ansatz that leads to the Freedman universe does not work and some new approach is required.

We may proceed in the following way:
First, instead of writing the spin connection in terms of elfbein using,
\ba
D_\mu{}^a{}_c e^c_\nu - D_\nu{}^a{}_c e^c_\mu = 0
\ea
with $D_\mu{}^a{}_c = \delta^a_c \partial_\mu + \frac{i}{2C} k_\mu{}^a{}_c (x) 
= \delta^a_c \partial_\mu + \omega_\mu{}^a{}_c (x)$.
We may eliminate $e_{\mu }^a$ from this equation using,
\ba
e_\mu^a = iD_\mu X^a
\ea
We get
\ba
\left[
\partial _{\mu }\omega^a{}_{\nu b}(x)-\partial _{\nu }\omega^a{}_{\mu b}(x)
+ \omega^a{}_{\mu d} \omega^d{}_{\nu b} (x)
- \omega^a{}_{\nu d} \omega^d{}_{\mu b} (x)
\right] X^d = 0
\ea
We solve this equation for $\omega_\nu{}^a{}_b$ in terms of $X^d(x)$. 
Then the equations (\ref{YYY}) and (\ref{YYYY}) will give the elfbein in terms
of spin connection oppositely to the usual process. 
The final process is to check whether the solution satisfy the ansatz 
$e_{\mu }^a = iD_{\mu } X^a$.
\end{comment}

\vspace{7pt}
These approaches are now under investigation,
and we hope that we can report on them soon.

%%%%%%%%%%%%%%%%%%%%%%%%%%%
\subsection*{Acknowledgments}
%%%%%%%%%%%%%%%%%%%%%%%%%%%

The authors would like to thank Professor S. Iso for introducing us to each other.
S.S. is partially supported by Grant-in-Aid for Scientific Research 
(No.\,16K17711) from Japan Society for the Promotion of Science (JSPS).
H.S. would like to thank Professor R. Peccei and Professor A. Kusenko 
for their hospitality at UCLA where part of this work was done.

\appendix

\vspace{14pt}
\noindent
{\large\bf Appendix}
\vspace{-7pt}

%%%%%%%%%%%%%%%%%%%%%%
\section{Notations}
\label{sec:ca}
%%%%%%%%%%%%%%%%%%%%%%

We consider the generators $G^A$ of the algebra $E_{11}\otimes\, \ell_1$.
The current is defined as
\ba
\Omega_\mu(x)  = J^A_\mu (x) G^A,
\ea
and the energy-momentum tensor is defined as
\ba
\Theta_{\mu\nu}(x) &=& \frac{1}{C}\left( \tr[\Omega_\mu \Omega_\nu] - \frac12 \eta_{\mu\nu} \tr[\Omega_\rho\Omega_\rho] \right)
\nt
&=& \frac{1}{C}\left( J^A_\mu(x) J^A_\nu(x) - \frac12 \eta_{\mu\nu} J^A_\rho(x) J^A_\rho(x) \right).
\ea
Here the generators are normalized so that $\tr[G^A G^B] = \delta^{AB}$ is satisfied.

In the nonlinear realization, we define the field $g(x)$ as 
\ba
g(x) = \exp[\phi^A(x) G^A],
\ea
and the current $\Omega_\mu$ can be written as
\ba
\Omega_\mu(x) 
&=& g^{-1}\partial_\mu g \nt
&=& \partial_\mu \phi^A(x) G^A 
+ \frac12 \partial_\mu \phi^A(x)\cdot \phi^B(x) [G^A, G^B]
+ \cdots.
\ea
% necessary for additional explanations for sec.1

For example, the brane charge currents $B_\mu^{abc}(x)$ and $B_\mu^{abcdef}(x)$
should be related to 
the 3-form field $C_{\mu\nu\rho}(x)$ and its dual 6-form field $C_{\mu\nu\rho\sigma\lambda\tau}(x)$ in 11d supergravity:
\ba
B_\mu^{abc}(x) &=& \partial_\mu C^{abc}(x) - 3 \partial_\mu k^{[a}{}_d(x) C^{bc]d}(x) +\cdots
\nt
B_\mu^{abcdef}(x) &=& \partial_\mu C^{abcdef}(x) + 6\partial_\mu k^{[a}{}_g (x) C^{bcdef]g} (x) +\cdots
\ea
where $C^{abc}(x) = e_\mu^a e_\nu^b e_\rho^c C^{\mu\nu\rho}(x)$ and 
$C^{abcdef}(x) = e_\mu^a e_\nu^b e_\rho^c e_\sigma^d e_\lambda^e e_\tau^f C^{\mu\nu\rho\sigma\lambda\tau}(x)$.

%Similarly, 
%the brane currents $W_\mu^{ab}(x)$ and $W_\mu^{abcde}(x)$ are also written as 
%\ba
%W_\mu^{ab}(x) &=& \partial_\mu \zeta^{ab}(x) + 2 \partial_\mu k^{[a}{}_c(x) \zeta^{b]c}(x) +\cdots \nt
%W_\mu^{abcde}(x) &=& \partial_\mu \zeta^{abcde}(x) -5 \partial_\mu k^{[a}{}_f (x)\zeta^{bcde]f} (x) +\cdots.
%\ea
%Here we define the tensors $\zeta^{ab}(x)$ and $\zeta^{abcde}(x)$.

%%%%%%%%%%%%%%%%%%%%%%%%%%%
\section{Commutation relations}
\label{sec:comm}
%%%%%%%%%%%%%%%%%%%%%%%%%%%

\subsection{Commutation relations among generators}

We list the commutation relations among all the generators in eq.\,(\ref{Omega}) below:
\ba
&&
[K_{ab}, K_{cd}] = 2 (-\eta_{b[c} K_{d]a} + \eta_{a[c} K_{d]b})
\nt &&
[K_{ab}, P_c] = -2\eta_{c[a} P_{b]}
\nt &&
[K_{ab}, S_{cde}] = 3(\eta_{b[c} S_{de]a} - \eta_{a[c} S_{de]b}) 
\nt &&
[K_{ab}, S_{cdefgh}] = 6(\eta_{a[c} S_{defgh]b} - \eta_{b[c} S_{defgh]a}) 
\nt &&
[K_{ab}, Z^{cd}] = 4 Z_{[a}{}^{[c} \delta^{d]}_{b]}
%[K^a{}_b, Z^{cd}] = 2Z^{a[c}\delta^{d]}_b + \frac12\delta^a_b Z^{cd} \nt
\nt &&
[K_{ab}, Z^{cdefg}] = -10  Z_{[a}{}^{[cdef} \delta^{g]}_{b]}
%[K^a{}_b, Z^{cdefg}] = 5Z^{a[cdef}\delta^{g]}_b + \frac12\delta^a_b Z^{cdefg} 
\ea
\ba
&&
[S_{abc}, P_d] = 3! \eta_{d[a} Z_{bc]} 
\nt &&
[S^{abc}, S_{def}] = -36 \delta^{[ab}_{[de} K^{c]}{}_{f]} + 2S^{abc}{}_{def}
% 72 -> 36 on 2017/9/20
\nt &&
[S_{abc}, S^{defghi}] = -4\cdot 5! \delta^{[def}_{abc} S^{ghi]} - 3S^{defghi}{}_{[ab,c]}
% \frac{2\cdot 6!}{3} = 4 \cdot 5!
\nt &&
[S_{abc}, Z^{de}] = 5! Z_{abc}{}^{de} - 6 \delta^{de}_{[ab} P_{c]}
\nt &&
[S_{abc}, Z^{defgh}] = \frac14 Z^{defgh}{}_{[ab,c]} - 2\delta^{[def}_{abc} Z^{gh]}
\ea
\ba
&&
[S_{abcdef}, P_g] = -6! \eta_{g[a} Z_{bcdef]} \nt %+ \cdots \nt
&&
[S_{abcdef}, Z^{gh}] =  -30 Z_{[abc}{}^{gh}{}_{de,f]}\nt
&&
[S_{abcdef}, Z^{defgh}] = 12 \delta_{[abcde}^{ghijk} P_{f]} - 84 \delta^{[ghi}_{[abc} Z_{def]}{}^{jk]}
\ea
\ba
[P_a, P_b] = [P_a, Z^{bc}] = [P_a, Z^{bcdef}]  = [Z^{ab}, Z^{cdefg}] = 0\,.
\ea
Here we note that
\ba
Z^{abcdefg,h} &:=& \epsilon^{abcdefg}{}_{ijkl} Z^{hijkl} \nt % \quad \text{(related to M5-brane)} \nt
S_{abcdefgh,i} &:=& R^{jklmnpqr,s}\eta_{aj}\ldots \eta_{is} - R_{abcdefgh,i}
\ea
where $R^{abcdefgh,i}$ and $R_{abcdefgh,i}$ are the generators of the $E_{11}$ algebra at level $3$ and $-3$, respectively. 
%(Does this mean we can set $S_{abcdef}{}^{[gh,i]} = 0$ ?)

\subsection{Commutation relations among currents}

We list some of the commutation relations among the currents (\ref{current_eom}) below:
\ba
%{}[k_0^{ab}(x), k_0^{cd}(y)]_{x_0=y_0} &=& 
% 2 (\eta^{a[c} k_0^{d]b} - \eta^{b[c} k_0^{d]a})(y)\delta(\vec x-\vec y)\nt
{}[k_0^{ab}(x), k_N^{cd}(y)]_{x_0=y_0} &=& 
 2 (\eta^{a[c} k_N^{d]b} - \eta^{b[c} k_N^{d]a}) (y)\delta(\vec x-\vec y) 
+ 2iC \eta^{ab,cd} \partial_N \delta(\vec x-\vec y) \nt
{}[k_0^{ab}(x), e_\mu^c(y)]_{x_0=y_0} &=& 
 -2 \eta^{c[a} e_\mu^{b]}(y)\delta(\vec x-\vec y) \nt
{}[k_0^{ab}(x), B_\mu^{cde}(y)]_{x_0=y_0} &=&
 3 (\eta^{b[c} B_\mu^{de]a} - \eta^{a[c} B_\mu^{de]b}) (y)\delta(\vec x-\vec y) \nt
{}[k_0^{ab}(x), B_\mu^{cdefgh}(y)]_{x_0=y_0} &=&
 6 (\eta^{a[c} B_\mu^{defgh]b} - \eta^{b[c} B_\mu^{defgh]a}) (y)\delta(\vec x-\vec y) \nt
{}[k_0^{ab}(x), W_\mu^{cd}(y)]_{x_0=y_0} &=&
 2 (\eta^{a[c} W_\mu^{d]b} - \eta^{b[c} W_\mu^{d]a}) \delta(\vec x-\vec y) \nt
{}[k_0^{ab}(x), W_\mu^{cdefg}(y)]_{x_0=y_0} &=&
 5 (\eta^{b[c} W_\mu^{defg]a} - \eta^{a[c} W_\mu^{defg]b}) \delta(\vec x-\vec y)
\ea
%\ba
%{}[e_0^a(x), e_0^b(y)]_{x_0=y_0} &=& 0 \nt
%{}[e_0^a(x), e_N^b(y)]_{x_0=y_0} &=&
% iC \eta^{ab} \partial_N \delta(\vec x-\vec y)
%\ea
\ba
{}[B_0^{abc}(x), e_\mu^d(y)]_{x_0=y_0} &=&
 6 \eta^{d[a} W_\mu^{bc]}(y)\delta(\vec x-\vec y) \nt
%
%{}[B_0^{abc}(x), B_{0,def}(y)]_{x_0=y_0} &=&
% (-18 \delta^{[ab}_{[de} k_0{}^{c]}{}_{f]}(y) + 2 B_0^{abc}{}_{def}(y)) \delta(\vec x-\vec y)\nt
{}[B_0^{abc}(x), B_{N,def}(y)]_{x_0=y_0} &=&
 (-36 \delta^{[ab}_{[de} k_N{}^{c]}{}_{f]}(y) + 2 B_N^{abc}{}_{def}(y)) \delta(\vec x-\vec y) 
 + 6iC\delta^{abc}_{def}\partial_N \delta(\vec x-\vec y)\nt
{}[B_0^{abc}(x), B_{N,defghi}(y)]_{x_0=y_0} &=&
(-4\cdot 5! \delta_{[def}^{abc} B_{N,ghi]} +\cdots ) \delta(\vec x-\vec y) \nt
{}[B_0^{abc}(x), W_{\mu, de}(y)]_{x_0=y_0} &=& 
 (5! W_\mu^{abc}{}_{de} - 6 \delta^{[ab}_{de} e_\mu^{c]})(y)\delta(\vec x-\vec y) \nt
{}[B_{0,abc}(x), W_{\mu}^{defgh}(y)]_{x_0=y_0} &=&
 (\frac14 W_\mu^{defgh}{}_{[ab,c]} - 2 \delta_{abc}^{[def} W_\mu^{gh]})(y)
 \delta(\vec x-\vec y) 
\ea
%
%\ba
%{}[W_0^{ab}(x), W_N^{cd}(y)]_{x_0=y_0} &=&
% \frac12 iC (\eta^{a[c}\eta^{d]b} - \eta^{b[c}\eta^{d]a}) \partial_N \delta(\vec x-\vec y) \nt
%{}[W_0^{abcde}(x), W_{N, fghij}(y)]_{x_0=y_0} &=&
% iC \delta^{abcde}_{defgh} \partial_N \delta(\vec x-\vec y) 
%\ea
%
\ba
{}[B_0^{abcdef}(x), e_\mu^g(y)]_{x_0=y_0} &=& 
 -6! \eta^{g[a} W_\mu^{bcdef]}(y) \delta(\vec x-\vec y) \nt
{}[B_0^{abcdef}(x), W_{\mu,gh}(y)]_{x_0=y_0} &=& 
 -30 \eps^{[abcde}{}_{ghijkl} W_\mu^{f]ijkl}(y) \delta(\vec x-\vec y) \nt
{}[B_{0,abcdef}(x), W_\mu^{ghijk}(y)]_{x_0=y_0} &=&
 (12 \delta_{[abcde}^{ghijk} e_{\mu,f]} - 84 \delta^{[ghi}_{[abc} W_{\mu,def]}{}^{jk]})
\delta(\vec x-\vec y) 
\ea
where $\eta^{ab,cd} := \frac12(\eta^{a[c}\eta^{d]b} - \eta^{b[c}\eta^{d]a})$
and $\delta^{abc}_{def} := \delta^{[a}_d \delta^b_e \delta^{c]}_f$.
The indices follow the same notations in the maintext: 
$\mu=0,\ldots,10$ and $N=1,\ldots,10$.


\begin{thebibliography}{99}

%\cite{Bergshoeff:1987cm}
\bibitem{Bergshoeff:1987cm}
  E.~Bergshoeff, E.~Sezgin and P.~K.~Townsend,
  ``Supermembranes and Eleven-Dimensional Supergravity,''
  Phys.\ Lett.\ B {\bf 189} (1987) 75.
%  doi:10.1016/0370-2693(87)91272-X
  %%CITATION = doi:10.1016/0370-2693(87)91272-X;%%
  %755 citations counted in INSPIRE as of 12 Aug 2017

%\cite{deWit:1988wri}
\bibitem{deWit:1988wri}
  B.~de Wit, J.~Hoppe and H.~Nicolai,
  ``On the Quantum Mechanics of Supermembranes,''
  Nucl.\ Phys.\ B {\bf 305} (1988) 545.
%  doi:10.1016/0550-3213(88)90116-2
  %%CITATION = doi:10.1016/0550-3213(88)90116-2;%%
  %723 citations counted in INSPIRE as of 12 Aug 2017

%\cite{Bagger:2007jr}
\bibitem{Bagger:2007jr}
  J.~Bagger and N.~Lambert,
  ``Gauge symmetry and supersymmetry of multiple M2-branes,''
  Phys.\ Rev.\ D {\bf 77} (2008) 065008
%  doi:10.1103/PhysRevD.77.065008
  [arXiv:0711.0955 [hep-th]].
  %%CITATION = doi:10.1103/PhysRevD.77.065008;%%
  %867 citations counted in INSPIRE as of 12 Aug 2017

%\cite{Gustavsson:2007vu}
\bibitem{Gustavsson:2007vu}
  A.~Gustavsson,
  ``Algebraic structures on parallel M2-branes,''
  Nucl.\ Phys.\ B {\bf 811} (2009) 66
%  doi:10.1016/j.nuclphysb.2008.11.014
  [arXiv:0709.1260 [hep-th]].
  %%CITATION = doi:10.1016/j.nuclphysb.2008.11.014;%%
  %805 citations counted in INSPIRE as of 12 Aug 2017

%\cite{Aharony:2008ug}
\bibitem{Aharony:2008ug}
  O.~Aharony, O.~Bergman, D.~L.~Jafferis and J.~Maldacena,
  ``$\cN=6$ superconformal Chern-Simons-matter theories, M2-branes and their gravity duals,''
  JHEP {\bf 0810} (2008) 091
%  doi:10.1088/1126-6708/2008/10/091
  [arXiv:0806.1218 [hep-th]].
  %%CITATION = doi:10.1088/1126-6708/2008/10/091;%%
  %1621 citations counted in INSPIRE as of 12 Aug 2017

%\cite{Kapustin:2009kz}
\bibitem{Kapustin:2009kz}
  A.~Kapustin, B.~Willett and I.~Yaakov,
  ``Exact Results for Wilson Loops in Superconformal Chern-Simons Theories with Matter,''
  JHEP {\bf 1003} (2010) 089
%  doi:10.1007/JHEP03(2010)089
  [arXiv:0909.4559 [hep-th]].
  %%CITATION = doi:10.1007/JHEP03(2010)089;%%
  %517 citations counted in INSPIRE as of 12 Aug 2017

%\cite{Drukker:2010nc}
\bibitem{Drukker:2010nc}
  N.~Drukker, M.~Marino and P.~Putrov,
  ``From weak to strong coupling in ABJM theory,''
  Commun.\ Math.\ Phys.\  {\bf 306} (2011) 511
%  doi:10.1007/s00220-011-1253-6
  [arXiv:1007.3837 [hep-th]].
  %%CITATION = doi:10.1007/s00220-011-1253-6;%%
  %295 citations counted in INSPIRE as of 12 Aug 2017

%\cite{Ho:2008ve}
\bibitem{Ho:2008ve}
  P.~M.~Ho, Y.~Imamura, Y.~Matsuo and S.~Shiba,
  ``M5-brane in three-form flux and multiple M2-branes,''
  JHEP {\bf 0808} (2008) 014
%  doi:10.1088/1126-6708/2008/08/014
  [arXiv:0805.2898 [hep-th]].
  %%CITATION = doi:10.1088/1126-6708/2008/08/014;%%
  %135 citations counted in INSPIRE as of 12 Aug 2017

%\cite{Chu:2008qv}
\bibitem{Chu:2008qv}
  C.~S.~Chu, P.~M.~Ho, Y.~Matsuo and S.~Shiba,
  ``Truncated Nambu-Poisson Bracket and Entropy Formula for Multiple Membranes,''
  JHEP {\bf 0808} (2008) 076
%  doi:10.1088/1126-6708/2008/08/076
  [arXiv:0807.0812 [hep-th]].
  %%CITATION = doi:10.1088/1126-6708/2008/08/076;%%
  %33 citations counted in INSPIRE as of 12 Aug 2017

%\cite{Ho:2009nk}
\bibitem{Ho:2009nk}
  P.~M.~Ho, Y.~Matsuo and S.~Shiba,
  ``Lorentzian Lie (3-)algebra and toroidal compactification of M/string theory,''
  JHEP {\bf 0903} (2009) 045
%  doi:10.1088/1126-6708/2009/03/045
  [arXiv:0901.2003 [hep-th]].
  %%CITATION = doi:10.1088/1126-6708/2009/03/045;%%
  %33 citations counted in INSPIRE as of 12 Aug 2017

%\cite{Kobo:2009gz}
\bibitem{Kobo:2009gz}
  T.~Kobo, Y.~Matsuo and S.~Shiba,
  ``Aspects of U-duality in BLG models with Lorentzian metric 3-algebras,''
  JHEP {\bf 0906} (2009) 053
%  doi:10.1088/1126-6708/2009/06/053
  [arXiv:0905.1445 [hep-th]].
  %%CITATION = doi:10.1088/1126-6708/2009/06/053;%%
  %8 citations counted in INSPIRE as of 12 Aug 2017

%\cite{Hanada:2012si}
\bibitem{Hanada:2012si}
  M.~Hanada, M.~Honda, Y.~Honma, J.~Nishimura, S.~Shiba and Y.~Yoshida,
  ``Numerical studies of the ABJM theory for arbitrary N at arbitrary coupling constant,''
  JHEP {\bf 1205} (2012) 121
%  doi:10.1007/JHEP05(2012)121
  [arXiv:1202.5300 [hep-th]].
  %%CITATION = doi:10.1007/JHEP05(2012)121;%%
  %71 citations counted in INSPIRE as of 12 Aug 2017

%\cite{Pasti:1997gx}
\bibitem{Pasti:1997gx}
  P.~Pasti, D.~P.~Sorokin and M.~Tonin,
  ``Covariant action for a $D = 11$ five-brane with the chiral field,''
  Phys.\ Lett.\ B {\bf 398} (1997) 41
%  doi:10.1016/S0370-2693(97)00188-3
  [hep-th/9701037].
  %%CITATION = doi:10.1016/S0370-2693(97)00188-3;%%
  %281 citations counted in INSPIRE as of 12 Aug 2017

%\cite{Lambert:2010wm}
\bibitem{Lambert:2010wm}
  N.~Lambert and C.~Papageorgakis,
  ``Nonabelian (2,0) Tensor Multiplets and 3-algebras,''
  JHEP {\bf 1008} (2010) 083
%  doi:10.1007/JHEP08(2010)083
  [arXiv:1007.2982 [hep-th]].
  %%CITATION = doi:10.1007/JHEP08(2010)083;%%
  %82 citations counted in INSPIRE as of 12 Aug 2017

%\cite{Honma:2011br}
\bibitem{Honma:2011br}
  Y.~Honma, M.~Ogawa and S.~Shiba,
  ``Dp-branes, NS5-branes and U-duality from nonabelian (2,0) theory with Lie 3-algebra,''
  JHEP {\bf 1104} (2011) 117
%  doi:10.1007/JHEP04(2011)117
  [arXiv:1103.1327 [hep-th]].
  %%CITATION = doi:10.1007/JHEP04(2011)117;%%
  %14 citations counted in INSPIRE as of 27 Sep 2017

%\cite{Ho:2011ni}
\bibitem{Ho:2011ni}
  P.~M.~Ho, K.~W.~Huang and Y.~Matsuo,
  ``A Non-Abelian Self-Dual Gauge Theory in $5+1$ Dimensions,''
  JHEP {\bf 1107} (2011) 021
%  doi:10.1007/JHEP07(2011)021
  [arXiv:1104.4040 [hep-th]].
  %%CITATION = doi:10.1007/JHEP07(2011)021;%%
  %56 citations counted in INSPIRE as of 27 Sep 2017

%\cite{Huang:2012tu}
\bibitem{Huang:2012tu}
  K.~W.~Huang,
  ``Non-Abelian Chiral 2-Form and M5-Branes,''
  arXiv:1206.3983 [hep-th].
  %%CITATION = ARXIV:1206.3983;%%
  %17 citations counted in INSPIRE as of 27 Sep 2017

%\cite{Samtleben:2011fj}
\bibitem{Samtleben:2011fj}
  H.~Samtleben, E.~Sezgin and R.~Wimmer,
  ``(1,0) superconformal models in six dimensions,''
  JHEP {\bf 1112} (2011) 062
%  doi:10.1007/JHEP12(2011)062
  [arXiv:1108.4060 [hep-th]].
  %%CITATION = doi:10.1007/JHEP12(2011)062;%%
  %84 citations counted in INSPIRE as of 12 Aug 2017

%\cite{Douglas:2010iu}
\bibitem{Douglas:2010iu}
  M.~R.~Douglas,
  ``On $D=5$ super Yang-Mills theory and (2,0) theory,''
  JHEP {\bf 1102} (2011) 011
%  doi:10.1007/JHEP02(2011)011
  [arXiv:1012.2880 [hep-th]].
  %%CITATION = doi:10.1007/JHEP02(2011)011;%%
  %161 citations counted in INSPIRE as of 12 Aug 2017

%\cite{Lambert:2010iw}
\bibitem{Lambert:2010iw}
  N.~Lambert, C.~Papageorgakis and M.~Schmidt-Sommerfeld,
  ``M5-Branes, D4-Branes and Quantum 5D super-Yang-Mills,''
  JHEP {\bf 1101} (2011) 083
%  doi:10.1007/JHEP01(2011)083
  [arXiv:1012.2882 [hep-th]].
  %%CITATION = doi:10.1007/JHEP01(2011)083;%%
  %158 citations counted in INSPIRE as of 12 Aug 2017

%\cite{West:2003fc}
\bibitem{West:2003fc} 
  P.~C.~West,
  ``$E_{11}$, SL(32) and central charges,''
  Phys.\ Lett.\ B {\bf 575}, 333 (2003)
%  doi:10.1016/j.physletb.2003.09.059
  [hep-th/0307098].
  %%CITATION = doi:10.1016/j.physletb.2003.09.059;%%

See also,
  P.~C.~West,
  %``E(11) and M theory,''
  Class.\ Quant.\ Grav.\  {\bf 18} (2001) 4443
%  doi:10.1088/0264-9381/18/21/305
  [hep-th/0104081],
  %%CITATION = doi:10.1088/0264-9381/18/21/305;%%
  %385 citations counted in INSPIRE as of 19 Sep 2017

  A.~G.~Tumanov and P.~West,
  %``E$_{11}$ must be a symmetry of strings and branes,''
  Phys.\ Lett.\ B {\bf 759} (2016) 663
%  doi:10.1016/j.physletb.2016.06.011
  [arXiv:1512.01644 [hep-th]],
  %%CITATION = doi:10.1016/j.physletb.2016.06.011;%%
  %10 citations counted in INSPIRE as of 19 Sep 2017

  A.~G.~Tumanov and P.~West,
  %``E11 in 11D,''
  Phys.\ Lett.\ B {\bf 758} (2016) 278
%  doi:10.1016/j.physletb.2016.04.058
  [arXiv:1601.03974 [hep-th]],
  %%CITATION = doi:10.1016/j.physletb.2016.04.058;%%
  %10 citations counted in INSPIRE as of 19 Sep 2017

  P.~West,
  %``A brief review of E theory,''
  Int.\ J.\ Mod.\ Phys.\ A {\bf 31} (2016) no.26,  1630043
%  doi:10.1142/S0217751X1630043X
  [arXiv:1609.06863 [hep-th]],
  %%CITATION = doi:10.1142/S0217751X1630043X;%%
  %5 citations counted in INSPIRE as of 19 Sep 2017

  D.~S.~Berman, H.~Godazgar, M.~J.~Perry and P.~West,
  ``Duality Invariant Actions and Generalised Geometry,''
  JHEP {\bf 1202} (2012) 108
%  doi:10.1007/JHEP02(2012)108
  [arXiv:1111.0459 [hep-th]].
  %%CITATION = doi:10.1007/JHEP02(2012)108;%%
  %110 citations counted in INSPIRE as of 07 Oct 2017

%\cite{Sugawara:2017fds}
\bibitem{Sugawara:2017fds}
  H.~Sugawara,
  ``Current Algebra Formulation of M-theory based on $E_{11}$ Kac-Moody Algebra,''
  Int.\ J.\ Mod.\ Phys.\ A {\bf 32} (2017) no.05,  1750024
%  doi:10.1142/S0217751X17500245
  [arXiv:1701.06894 [hep-th]].
  %%CITATION = doi:10.1142/S0217751X17500245;%%

%\cite{Englert:2008ft}
\bibitem{Englert:2008ft}
  F.~Englert and L.~Houart,
  ``The Emergence of fermions and the $E_{11}$ content,''
%  doi:10.1007/978-0-387-87499-9_9
  arXiv:0806.4780 [hep-th].
  %%CITATION = doi:10.1007/978-0-387-87499-9_9;%%
  %1 citations counted in INSPIRE as of 21 Sep 2017

%\cite{Berman:2014hna}
\bibitem{Berman:2014hna}
  D.~S.~Berman and F.~J.~Rudolph,
  ``Strings, Branes and the Self-dual Solutions of Exceptional Field Theory,''
  JHEP {\bf 1505} (2015) 130
%  doi:10.1007/JHEP05(2015)130
  [arXiv:1412.2768 [hep-th]].
  %%CITATION = doi:10.1007/JHEP05(2015)130;%%
  %25 citations counted in INSPIRE as of 07 Oct 2017

%\cite{Sugawara:1967rw}
\bibitem{Sugawara:1967rw}
  H.~Sugawara,
  ``A Field theory of currents,''
  Phys.\ Rev.\  {\bf 170} (1968) 1659.
%  doi:10.1103/PhysRev.170.1659
  %%CITATION = doi:10.1103/PhysRev.170.1659;%%
  %371 citations counted in INSPIRE as of 05 Sep 2017

%\cite{Pope:1989cr}
\bibitem{Pope:1989cr}
  C.~N.~Pope and K.~S.~Stelle,
  ``SU($\infty$), SU${}_+$($\infty$) and Area Preserving Algebras,''
  Phys.\ Lett.\ B {\bf 226} (1989) 257.
%  doi:10.1016/0370-2693(89)91191-X
  %%CITATION = doi:10.1016/0370-2693(89)91191-X;%%
  %44 citations counted in INSPIRE as of 19 Sep 2017

%\cite{Frappat:1989gn}
\bibitem{Frappat:1989gn}
  L.~Frappat, E.~Ragoucy, P.~Sorba, F.~Thuillier and H.~Hogaasen,
  ``Generalized Kac-Moody Algebras and the Diffeomorphism Group of a Closed Surface,''
  Nucl.\ Phys.\ B {\bf 334} (1990) 250.
%  doi:10.1016/0550-3213(90)90663-X
  %%CITATION = doi:10.1016/0550-3213(90)90663-X;%%
  %9 citations counted in INSPIRE as of 19 Sep 2017

D.~Persson and N.~Tabti,
``Lectures on Kac-Moody Algebras with Applications in (Super-)Gravity,''
a lecture note uploaded to
http://www.ulb.ac.be/sciences/ptm/pmif/Rencontres/KMModaveLectures2007.pdf

%\cite{Brink:2008hv}
\bibitem{Brink:2008hv}
  L.~Brink, S.~S.~Kim and P.~Ramond,
  ``$E_{8(8)}$ in Light Cone Superspace,''
  JHEP {\bf 0807} (2008) 113
%  doi:10.1088/1126-6708/2008/07/113
  [arXiv:0804.4300 [hep-th]].
  %%CITATION = doi:10.1088/1126-6708/2008/07/113;%%
  %6 citations counted in INSPIRE as of 19 Sep 2017

%\cite{Bardakci:1968zz}
\bibitem{Bardakci:1968zz}
  K.~Bardakci, Y.~Frishman and M.~B.~Halpern,
  ``Structure and Extensions of a Theory of Currents,''
  Phys.\ Rev.\  {\bf 170} (1968) 1353.
%  doi:10.1103/PhysRev.170.1353
  %%CITATION = doi:10.1103/PhysRev.170.1353;%%
  %40 citations counted in INSPIRE as of 05 Sep 2017

%\cite{Tseytlin:1996bh}
\bibitem{Tseytlin:1996bh}
  A.~A.~Tseytlin,
  ``Harmonic superpositions of M-branes,''
  Nucl.\ Phys.\ B {\bf 475} (1996) 149
%  doi:10.1016/0550-3213(96)00328-8
  [hep-th/9604035].
  %%CITATION = doi:10.1016/0550-3213(96)00328-8;%%
  %366 citations counted in INSPIRE as of 20 Sep 2017

\end{thebibliography}
\end{document}